\documentclass[aps,pra,twocolumn,groupedaddress,showpacs,floatfix]{revtex4}

\usepackage{amsmath,mathrsfs,graphics,epsfig,ulem,longtable}
\usepackage[usenames]{color}

\newcommand{\bd}[1]{\mathbf{#1}}

\newcommand{\partd}[2]{\frac{\partial{#1}}{\partial{#2}}}

\newcommand{\ud}{\,\mathrm{d}}

\begin{document}


\title{Pseudospectral calculation of the wave function of helium and the negative hydrogen ion}


\author{Paul E. Grabowski}
\email[]{peg9@cornell.edu}
\affiliation{Department of Physics, Cornell University, Ithaca, New York 14853, USA}
\author{David F. Chernoff}
\email[]{chernoff@astro.cornell.edu}
\affiliation{Department of Astronomy, Cornell University, Ithaca, New York 14853, USA}


\date{\today}

\begin{abstract}
We study the numerical solution of the nonrelativistic
Schr\"{o}dinger equation for two-electron atoms in ground and excited
{\it S} states using pseudospectral (PS) methods of calculation. 
The calculation achieves convergence rates for the energy, Cauchy
error in the wave function, and variance in local energy that are
exponentially fast for all practical purposes. The method
requires three separate subdomains to handle the
wave function's cusplike behavior near the two-particle
coalescences. The use of three subdomains is essential to 
maintaining exponential convergence and is more computationally
efficient than a single subdomain. 
A comparison of several different treatments of the
cusps suggests that the simplest prescription is sufficient.
We investigate two alternate methods for handling the semi-infinite
domain, one which involves a sequence of truncated versions of the
domain and the other which employs an algebraic mapping of the
semi-infinite domain to a finite one and imposes no explicit cutoffs
on the wave function. The latter prescription proves superior.
For many purposes it proves unnecessary to handle the three-particle
coalescence in a special way. The presence of logarithmic terms in the
exact solution is expected to limit the convergence to being
nonexponential but the only clear evidence of that is the rate of
convergence of derivatives near the three-particle coalescence point.
Higher resolution than achieved in this work will ultimately be needed
to see its limiting effect on other measures of error.
As developed and applied here the PS method has many virtues: no
explicit assumptions need be made about the asymptotic behavior of the
wave function near cusps or at large distances, the local energy
(${\cal{H}}\psi/\psi$) is exactly equal to the calculated global
energy at all collocation points, local errors go down everywhere with
increasing resolution, the effective basis using Chebyshev polynomials
is complete and simple, and the method is easily extensible to other bound
states. As the number of collocation points grows, the method achieves
exponential convergence up to the resolution tested.
This study
serves as a proof-of-principle of the method for more general two- and possibly
three-electron applications.
\end{abstract}


\maketitle

\section{Introduction}

The nonrelativistic, two-electron atom (H$^-$, He, Li$^+$) is the
simplest ``hard'' problem in quantum mechanics. It involves strong
electron-electron correlations, nontrivial symmetry considerations,
and single as well as double continua.  Many
different solution techniques have been developed and applied over the
past 80 years. A thorough understanding of this simple system is
important not only because of its direct relevance to experimental studies in
atomic physics but also because the best methods of solution may suggest
generalizations applicable to multielectron and/or multiatom systems.

Our own interest in this problem arose from investigating bound-free
and free-free opacity of the negative hydrogen ion H$^-$.  As first
conjectured by Wildt \cite{Wildt1939}, H$^-$ gives the greatest
contribution to opacity in the atmosphere of the Sun and many other
stars. The photo-absorption cross section of H$^-$ is known to an
accuracy of a few percentage points \cite{John1988} but little attention
has been devoted to H$^-$ in less-than-ideal circumstances (high
density, high magnetic field, etc.) of relevance to astrophysical
applications.  We sought a first-principles approach that would allow
``exact'' calculations of initial and final states as part of these
investigations and were led to reconsider this classic problem.

Ideally, there would exist a simple method capable of handling any
two-electron state in the presence of a nucleus with any angular
momentum whether bound or free. In practice, many individual methods
have been formulated each having somewhat more specific goals. A common
starting point, for example, is finding the ground-state energy for
zero total angular momentum. 

It is not possible in a single article, let alone an introduction, to
review the full range of methods that have been developed and
explored.  We can briefly compare the strengths and weaknesses of a
few select approaches by assessing each in terms of the generality (is
it applicable to all states or just the ground state?), the capability
of achieving an exact solution of the nonrelativistic Schr\"{o}dinger
equation in the limit of infinite nuclear mass (is it {\it in
  principle} capable of finding an exact solution [in the
  aforementioned sense] or are there intrinsic approximations?), the
degree of tuning required (is it straightforward to apply or does it
require an enlightened guess for, say, the choice of basis functions?)
and, of course, the computational effort for a given level of
accuracy.

The asymptotic
rate of convergence of some error $R_n$ as a function of the number $n$ of 
basis functions, grid size, etc. is of central importance in
evaluating a numerical method.
To characterize the convergence rate the definitions of Boyd \cite{Boyd2000}
are used in this article
and reproduced here. The algebraic index of convergence, $k$ is
defined as the maximum $k$ so that
\begin{equation}
\lim_{n\rightarrow\infty} |R_n| n^k <\infty.
\end{equation}
If $k$ is finite then $R_n$ converges algebraically. 
The simplest example of algebraic convergence is an error $R_n\propto 1/n^k$.
If $k$ is 
infinite then $R_n$ converges exponentially. This latter category
is subdivided into three cases defined by the value of 
\begin{equation}
l=\lim_{n\rightarrow\infty} \frac{\log{|R_n|}}{n}.
\end{equation}
If $l$ is zero, a finite positive number, or infinite, the rate
is subgeometric, geometric, or supergeometric, respectively.
For example, if the error $R_n\propto \exp(-n^m)$, the conditions $0<m<1$, $m=1$, and $m>1$
correspond to subgeometric, geometric, and supergeometric convergence, respectively.

\subsection{Variational method for two electrons and nucleus: ground state}

The first numerical explorations of two-electron ground states
adopted the approach of minimizing the global
energy. Once Hylleraas determined that
only three coordinates were needed to represent the wave function for
{\it S} states he carried out such variational calculations (prior to the
advent of computers) \cite{Hylleraas1929}. Pekeris and coworkers
\cite{Pekeris1958,Pekeris1962,FrankowskiPekeris1966,AccadEtAl1971}
did the first high precision calculations on computers, expanding the
wave function in terms of Laguerre polynomials of linear combinations
of the interparticle distances times an appropriate exponential falloff. They determined
the energy of H$^-$ to eight decimal places and that of He to nine,
calculations that were the gold standard for several decades.

Variational methods have been highly successful at calculating
extremely precise eigenvalues of the ground state of two-electron
atoms. Indeed, eigenvalue energies have been calculated to
numerical accuracy---at least 42 digits---that {\it far} exceeds the accuracy
of the underlying physical description based on nonrelativistic
equations of motion
\cite{BakerEtAl1990,Drake1999,Korobov2000,Frolov2000,Korobov2002,DrakeEtAl2002,Frolov2006,
  Frolov2007,NakashimaNakatsuji2007,NakashimaNakatsuji2008,KurokawaEtAl2008}.
A clear strength of the general variational approach is that intrinsic
approximations to the Hamiltonian operator need not be made. The principle
drawbacks are related to the difficulties inherent in the
selection of the basis: it should be complete so that convergence to the
exact solution is possible and efficient so that finite numbers of
elements do a good job representing the wave function.  

Significant progress in choice of the basis for the two-electron
problem has taken place.  The inclusion of new functions
(e.g., logarithmic terms) typically motivated by known limiting forms
of the wave function improves the rate of convergence
\cite{Kinoshita1957,Schwartz1960,Schwartz1962,FrankowskiPekeris1966,
BakerEtAl1990,Korobov2000,ThakkarKoga2003,Forrey2004}. Furthermore,
without special additions, some bases are simply incapable of
representing the exact solution \cite{BartlettEtAl1935,Bartlett1937}.
Klahn and Morgan have shown that there are examples where the 
expectation value of an operator (i.e., $r^k$ with $k\ge 6$)
converges to the incorrect value or diverges even if the basis is
complete. In their example the basis cannot accurately represent the 
derivatives of the hydrogenic solution at $r=0$ 
\cite{KlahnMorgan1984}. When employing such bases, one must always check
that the physical property one is calculating is converging properly.

Schwartz \cite{Schwartz2006} surveyed the convergence rate of the
error in the ground-state energy eigenvalue achieved by many different
strategies for basis set selection.  His results for the error may be
expressed as a function of $n$, the total number of basis functions
selected according to a well-defined procedure. The error generally
converges algebraically with index, $1.5 \le k \le 8.3$ ($\propto n^{-k}$),
depending on the basis. The range in $k$ highlights the significance
that a good choice of basis can have on the asymptotic convergence of a
calculation. One basis set, which included a single power of a
logarithm, appeared to converge exponentially fast as $\sigma^{-n}$
with $\sigma$ in the range $0.51$-$0.54$. Such exponential behavior
is often assumed of variational methods if the basis can accurately describe
the behavior of the wave function everywhere. That is, the basis includes
functions which have the same analytic and nonanalytic behavior as the exact solution.

Loosely speaking,
even when convergence is assured, the accuracy of the variationally
inferred wave function (by many different measures) is much
less than that of the energy eigenvalue.  Parts of the wave function
that have a small effect on the total energy are not well-constrained
by lowering the energy.  An alternative strategy to minimizing the
global energy is to minimize the variance in the local energy instead
\cite{Bartlett1955,Coldwell1977,UmrigarEtAl1988}. This approach can
produce better local values of the wave function but leads to
nonlinear minimization problems which are more difficult to handle
numerically (minimization of the variance in local energy with 
respect to parameters in the trial wave function) but still tractable
because one need not calculate the global energy at each step. 

\subsection{Variational method for excited states}

Variational methods \cite{Pekeris1962,AccadEtAl1971,KonoHattori1984,KonoHattori1985,Drake1999}
have been successful at calculating precise excitation energies. 
In general, variational
methods extend naturally to excited, bound states whenever the variational parameters
enter in a linear fashion. It is then straightforward to find multiple
eigenstates of the linear system. The more highly excited the state the 
less converged the energy is, especially if the basis was optimized in order to 
reproduce the features of the ground state only 
\footnote{Drake \cite{Drake1999} points out that 
alternative methods can calculate the singly excited spectrum. For high angular momentum,
the system can be treated as an electron in the field of a perturbed core with 
higher and higher moments of the core being included for higher and higher accuracy. 
This calculation can be done analytically for the two-electron problem.
The approximation here is that the electron correlation energy is ignored, but
this is very small for such states.
For large principle quantum numbers, quantum defect approximations work well.
The energy is proportional to $1/(n-\delta_l)^2$, instead of the usual
$1/n^2$, where $n$ is the principle quantum number of the excited electron,
$l$ is its angular momentum, and $\delta_l$ gives the effect of the screening 
due to the inner electron. 
}.



\subsection{Fourier spectral expansion}

Griebel and Hamaekers \cite{GriebelHamaekers2007} developed a Fourier
expansion method for multidimensional quantum mechanical systems.
They apply the hyperbolic cross truncation to their basis and show 
that for smooth solutions the exponential convergence rate is not dependent
on the dimensionality of the system. They calculate the energies of several
different systems including hydrogen and helium. Unfortunately, they fail 
to achieve exponential convergence because the cusps were not properly treated,
and hence their highest resolution runs for hydrogen and helium are only good to
about 2 and 10\%, respectively.

\subsection{Specialized methods for two-electron systems}

Haftel and Mandelzweig and later other collaborators
\cite{HaftelMandelzweig1983,HaftelMandelzweig1987,HaftelMandelzweig1988,
HaftelMandelzweig1989,HaftelMandelzweig1990,KrivecEtAl1991,KrivecEtAl2000}
have presented an exact treatment of two-electron atoms that begins by
factoring out the correct cusp behavior and posing the problem in
terms of the remaining part of the wave function. This piece which is
continuous up to first derivatives is expanded in terms of
hyperspherical harmonics yielding a set of coupled ordinary
differential equations for coefficients which are 
functions of the hyperradius. The method fully accounts for the asymptotic behavior near
coalescence points and yields results with energies good to one part
in $10^9$.  The hyperspherical harmonic expansion converges more
quickly than if the cusp is not explicitly accommodated for, 
but remains algebraic because of the higher-order discontinuities.
The method accurately determines bound excited states as well.

\subsection{Direct solution of partial differential equation for bound and continuum states}

Most of the bound-state techniques mentioned thus far are unsuitable
for calculating continuum-states. In fact, continuum state
calculations rely on totally different variational methods. The main
ones are {\it R}-matrix \cite{BurkeBerrington1993}, Schwinger variational
\cite{Huo1995}, and the complex Kohn variational
\cite{RescignoEtAl1995} methods.  Accuracy for these methods lags far
behind that of bound-state calculations.

Roughly speaking, the source of some of these difficulties is
related to describing the wave function over an infinite volume
while simultaneously controlling the errors of greatest significance.
Typically, linear variational methods are equivalent to spectral
expansions of the wave function. The control one has over the accuracy
of an approximate description of the wave function is indirect
via the choice of the expansion.
Instead, one may be motivated for both bound and continuum problems to
consider solving the partial differential equation directly on a grid
where a greater degree of local control is possible.

Finite difference
methods (FDM)
\cite{WinterEtAl1968,BarracloughMooney1971,HawkHardcastle1979,EdvardssonEtAl2005}
and Finite element methods (FEM)
\cite{LevinShertzer1985,BraunEtAl1993,AckermannShertzer1996,SchweizerEtAl1999,ZhengYing2004a,ZhengYing2004b}
represent the solution and the differential equation on a discrete
grid. The FDM grid is usually evenly spaced with derivatives
calculated to some small (usually second) order. FEM uses subdomains,
concentrating grid points where more accuracy is needed.  Recent work
achieves as many as seven decimal places in the energy of the ground
state but produces surprisingly nonsmooth wave functions
\cite{ZhengYing2004a,ZhengYing2004b}. The rate of convergence of these methods
is limited by the order of the representation of derivatives and is always 
algebraic with some small index dependent on the order used for derivatives.


\subsection{Pseudospectral approach}

Some of the above considerations motivate an investigation of the
pseudospectral (PS) method. Like FDM and FEM methods the PS method
represents the wave function by values on a discrete grid of points
rather than by coefficients of a spectral expansion. 
However, the points are selected in a different
manner and the derivative order increases with grid resolution. 
Roots of Jacobi polynomials are chosen in order to make the asymptotic 
rate of convergence of an analytic function constant
across the entire finite nonperiodic domain. 
Such a choice also has the advantage that exponential convergence can 
be lost only by nonanalytic behavior
within the domain. By contrast, an equispaced 
grid is sensitive to singularities nearby in the complex plane and can lead to 
divergences when interpolating near the endpoints (Runge phenomenon).
Of all the Jacobi polynomials, Chebyshev polynomials vary the least over
$[-1,1]$ and hence produce the smallest
residual from the PS method. The mathematical theory of nonsmooth functions
is not well developed and precise convergence rates are usually calculated
empirically \cite{Fornberg1996}. 

The PS method \cite{Fornberg1996,Boyd2000,NumericalRecipes} has seen successes 
in many fields including fluid dynamics \cite{CanutoEtAl1988}, relativistic astrophysics
\cite{BonazzolaEtAl1999}, and numerical relativity \cite{KidderFinn2000,PfeifferEtAl2003}.
When the underlying solution is smooth, the PS method typically requires
less computational run time and less memory than FDM and FEM
to achieve comparable precision.
The method has been applied in quantum
mechanics to solve the full Schr\"{o}dinger equation for a single electron
\cite{FattalEtAl1996,Borisov2001,BoydEtAl2003}. In addition, various simplifications of the multielectron 
Schr\"{o}dinger equation have been treated, including the Hartree-Fock approximation \cite{Friesner1985,Friesner1986,Friesner1987,RingnaldaEtAl1990,GreeleyEtAl1994,HeylThirumalai2009}, 
M{\o}ller-Plesset perturbation theory \cite{MurphyEtAl1995}, and density functional theory 
\cite{MurphyEtAl2000,KoEtAl2008}.

To the authors' knowledge, no one has solved the full
three-dimensional Schr\"{o}dinger equation for heliumlike systems
(the ``exact'' problem) using PS methods.  This article implements the
method, investigates several design choices and calculates ground and
excited bound {\it S} states.  The convergence rate is used as the metric to
characterize different grid choices, alternative methods for handling
regularity conditions and other practical considerations needed for an
efficient algorithm. No attempt to reproduce the ultrahigh precision results of variational
methods is made. The calculation employs a standard Chebyshev basis without any specialized tuning. The eigenvalue and eigenfunction problems are solved by a standard method. All calculations are done on a single processor with a speed of 6 GHz and 8 GB of memory.

The PS method is expected to be supergeometric on a finite computational domain
if no singularities exist in the solution anywhere in the complex plane,
geometric if singularities are only outside the domain, and algebraic
if singularities exist within domain. If the domain is infinite or 
semi-infinite, subgeometric convergence is expected when no singularities
are in the domain, and algebraic convergence is expected otherwise \cite{Boyd2000}.

\section{Setting Up the Problem}

Let $\bd{z_i}$ and $\nabla_i^2$ be the position vector and the
Laplacian of the coordinates, respectively, of the $i$th
electron if $i$ is $1$ or $2$ and of the nucleus if $i$ is $3$.  The
nonrelativistic Schr\"{o}dinger equation for a heliumlike system is
\begin{equation}
{\cal{H}}=-\frac{1}{2}\left(\frac{\nabla_1^2}{m}+\frac{\nabla_2^2}{m}+\frac{\nabla_3^2}{M}\right)+\cal{V},
\end{equation}
where
\begin{equation}
{\cal{V}}=-\frac{Z}{|\bd{z_1}-\bd{z_3}|}-\frac{Z}{|\bd{z_2}-\bd{z_3}|}+\frac{\alpha}{|\bd{z_1}-\bd{z_2}|},
\end{equation}
$Z$ is the nuclear charge, $\alpha=1$ unless the
electron-electron interaction is suppressed ($\alpha=0$), $m$ is the
mass of the electron and $M$ is the mass of the nucleus.  The units
are $e=\hbar=1/4\pi\epsilon_0=1$. The Hamiltonian acts on functions of
nine dimensions, i.e., three coordinate positions for each particle.

The relative and center of mass coordinates are
\begin{eqnarray}
\bd{r_1}&=&\bd{z_1}-\bd{z_3}\\
\bd{r_2}&=&\bd{z_2}-\bd{z_3}\\
\bd{R}&=&\frac{m(\bd{z_1}+\bd{z_2})+M \bd{z_3}}{M+2m} .
\end{eqnarray}
Define the coordinates
\begin{eqnarray}
r_1&=&|\bd{r_1}|\\
r_2&=&|\bd{r_2}|\\
r_{12}&=&|\bd{r_1}-\bd{r_2}|,
\end{eqnarray}
and rewrite the Hamiltonian
\begin{equation}
{\cal{H}}={\cal{T}}_0+{\cal{T}}_{cm}+{\cal{T}}_{mp}+{\cal{V}},
\end{equation}
where
\begin{eqnarray}
{\cal{T}}_0&=&-\frac{1}{2\mu}(\nabla_{r_1}^2+\nabla_{r_2}^2),\\
{\cal{T}}_{cm}&=&-\frac{1}{2(M+2m)}\nabla_R^2,\\
{\cal{T}}_{mp}&=&-\frac{1}{M}\bd{\nabla_{r_1}}\cdot \bd{\nabla_{r_2}},\\
{\cal{V}}&=&-\frac{Z}{r_1}-\frac{Z}{r_2}+\frac{\alpha}{r_{12}},
\end{eqnarray}
$\mu=mM/(M+m)$ is the reduced mass of the electron and nucleus,
and $\bd{\nabla_x}$ is the gradient operator with respect to the
vector $\bd{x}$.

In the center-of-mass frame ${\cal{T}}_{cm}$ may be dropped bringing
to six the number of nontrivial coordinates on which the
wave function depends.  Because $m \ll M$, the mass polarization term
${\cal{T}}_{mp}$ is often ignored or treated perturbatively.  While unnecessary for many
methods including PS, we use the infinite nuclear mass approximation ($M=\infty$) to facilitate
comparison with previous results.  In units with $m=1$ (atomic units) the
Hamiltonian is
\begin{equation}
{\cal{H}}=-\frac{1}{2}(\nabla_{r_1}^2+\nabla_{r_2}^2)-\frac{Z}{r_1}-\frac{Z}{r_2}+\frac{\alpha}{r_{12}}.
\end{equation}
Atomic units are used throughout the rest of this article.

This operator is elliptic.  All boundaries in physical space require
specification of the function or its normal derivative or some
combination of the two \cite{Arfken1970}.
In the ideal problem, the physical boundary is at infinity where the
wave function must be zero.
%
%
The existence of the Coulomb potential's singular points at $r_1=0$,
$r_2=0$, and $r_{12}=0$ introduces complications in any formal and practical
analysis.  Before the exact nature of the Hermitian Hamiltonian
operator and its spectrum was understood, Kato \cite{Kato1951b} showed
that discrete eigenstates existed for the specific case of helium.  In
later work Kato \cite{Kato1951a} showed that the wave function must be
finite at the singular points (which is also true everywhere else),
and that the first derivative of the wave function on the domain
excluding the singular points is bounded. This result allows
discontinuities in the first derivative at the singular points, called
Kato cusps. Generally, higher derivatives are not bounded at the
singular points.

In any numerical treatment of the Hamiltonian operator a decision must
be made about how to handle the singular points. In a formal mathematical 
sense, quantities at the singularities are well defined only in the limit as
one approaches the singularity. This creates an effective inner boundary
about such points on which additional conditions on the function and its
normal derivative may be specified. Such conditions are exploited to
guarantee regularity in the limit that the excised region shrinks to a
point. This article assumes
that it is correct to excise such a point, either explicitly or implicitly. 
%

\section{Coordinates and the Hamiltonian}
The heliumlike atom is made of three particles: a nucleus and two
electrons.  Six coordinates are required to describe relative
positions. Three coordinates describe the precise shape and size of
the triangle with a particle at each vertex, and the other three
describe the orientation of that triangle in space (often taken to be
Euler angles).  The wave function for {\it S} states is completely
independent of the latter three \cite{Hylleraas1929}. For nonzero
angular momentum one first expands the wavefunction in generalized
spherical harmonics of the Euler angles. Only a finite number of terms
are needed for a given total angular momentum and its $z$ component,
and the Shr\"{o}dinger equation becomes a finite set of coupled partial
differential equations for the remaining three variables
(e.g., Refs. \cite{BhatiaTemkin1964,BhatiaTemkin1965}).

Two useful sets of coordinates for the
triangle are \mbox{$\{r_1, r_2, r_{12}\}$}
and \mbox{$\{r_1, r_2,
\theta_{12}\}$}, where $r_1$ and $r_2$ are the proton-electron
distances, $r_{12}$ is the electron-electron distance, and
$\theta_{12}$ is the angle between the vectors pointing to the two
electrons. Four additional useful sets of coordinates \mbox{$\{\rho
\textrm{ or } x, \phi, C\}$} and \mbox{$\{\rho \textrm{ or } x, \zeta, B\}$}
are defined by
\begin{eqnarray}
r_1 & = & \rho \cos{\phi}\\
r_2 & = & \rho \sin{\phi}\\
C & = & -\cos{\theta_{12}}\\
\sqrt{2}\sin{\zeta} & = & \sqrt{1+C \sin{2\phi}}\\
B & = &\frac{\cos{2\phi}}{\sqrt{1-C^2\sin^2{2\phi}}}\\
x & = &\frac{1-\rho}{1+\rho}.
\end{eqnarray}
The ranges of these variables are given by:
\begin{equation}
\begin{array}{c}
0\le r_1,r_2,\rho < \infty\\ 
|r_1-r_2|\le r_{12} \le r_1+r_2\\
0\le \theta_{12} \le \pi\\
0\le \phi,\zeta\le \pi/2 \\
-1\le x,C,B\le 1.
\end{array}
\end{equation}
The coordinate $x$ maps the semi-infinite domain to a finite
domain \footnote{One can introduce a free parameter $L$ by defining
  $x=(1-\rho/L)/(1+\rho/L)$ and vary $L$ to optimize convergence but
  doing this led to only slight improvements. This finding is in agreement
  with Boyd et al. \cite{BoydEtAl2003}, who
  showed that $L$ has a small effect for the hydrogen atom when using a Chebyshev
  basis.}. This simple choice
works well because the wave function is exponentially small at large
$\rho$ for bound states which are the topic of interest here.

After integrating over the Euler angles the volume elements are
\begin{equation}
\int d^3 \bd{r_1}\, d^3 \bd{r_2}=2\pi^2\left\{ \begin{array}{l}
4 \int r_1 r_2 r_{12} dr_1\, dr_2 \,dr_{12}\\
4 \int r_1^2 r_2^2 \sin{\theta_{12}} dr_1\, dr_2\, d\theta_{12}\\
\int \rho^5 \sin^2{2\phi}d\rho\, d\phi\, dC\\
\int \rho^5 \sin^2{2\zeta}d\rho\, d\zeta\, dB\\
2 \int \frac{(1-x)^5}{(1+x)^7} \sin^2{2\phi}dx\, d\phi\, dC\\
2 \int \frac{(1-x)^5}{(1+x)^7} \sin^2{2\zeta}dx\, d\zeta\, dB.
\end{array}\right.
\end{equation} 

The Hamiltonian for {\it S} states can be written in hyperspherical coordinates as:
\begin{eqnarray}
\label{Hamiltonian}
{\cal{H}}&=& {\cal{T}}_\rho + \rho^{-2}({\cal{T}}_\phi + \csc^2{2\phi}{\cal{T}}_C) + \rho^{-1}{\cal{U}}\\ 
&=&{\cal{T}}_\rho + \rho^{-2}({\cal{T}}_\zeta + \csc^2{2\zeta}{\cal{T}}_B) + \rho^{-1}{\cal{U}} \mbox{,}
\end{eqnarray}
where
\begin{eqnarray}
{\cal{T}}_\rho&=&-\frac{1}{2}\partial_{\rho\rho}-\frac{5}{2\rho}\partial_\rho\\
&=&-\frac{(1+x)^4}{8}\partial_{xx}+\frac{(1+x)^3(4+x)}{4(1-x)}\partial_x\\
{\cal{T}}_\phi&=&-\left(\frac{1}{2}\partial_{\phi\phi}+2\cot{2\phi}\partial_\phi\right)\\
{\cal{T}}_C&=&-2\left((1-C^2)\partial_{CC}-2C\partial_C\right)\\
{\cal{T}}_\zeta&=&-\left(\frac{1}{2}\partial_{\zeta\zeta}+2\cot{2\zeta}\partial_\zeta\right)\\
{\cal{T}}_B&=&-2\left((1-B^2)\partial_{BB}-2B\partial_B\right)\\
{\cal{U}}&=&\frac{\alpha }{\sigma[C,\phi]}-Z\csc{\phi}-Z\sec{\phi}\\
&=&\frac{\alpha}{\sqrt{2}\sin{\zeta}}-\frac{Z\sqrt{2}}{\sigma[B,\zeta]}
-\frac{Z\sqrt{2}}{\sigma[-B,\zeta]},
\end{eqnarray}
and
\begin{equation}
\sigma[x,y]=\sqrt{1+x\sin{2y}}.
\end{equation}

\section{\label{Singularities}The Singular Points in the Hamiltonian}

PS methods are very sensitive to discontinuous
derivatives of any order.  If such discontinuities exist, the method
loses its exponential convergence and artificial oscillations may
occur. The wave function has discontinuities only at the singular
points which thus require special attention. Myers {\it et al} \cite{MyersEtAl1991} 
discuss these singularities in detail. Here we reproduce some of their discussion 
for completeness.

\subsection{Two-particle coalescences}

There exist three lines corresponding to two-particle coalescences: two
for the proton and each electron at $\phi=0$ and $\phi=\pi/2$ and one for
the two electrons at $\zeta=0$.  Only one of the proton-electron
coalescence lines need appear in the numerical domain which
takes advantage of the explicit symmetry of the spatial part of the
wave function about $\phi=\pi/4$.

Kato \cite{Kato1957} analyzed the discontinuity in the derivative of a
wave function at two particle coalescence points and showed that
\begin{equation}
\label{KatoCuspCond}
\left.\frac{\partial\hat{\psi}}{\partial r}\right|_{r=0}=\mu_{ij}q_i q_j\psi(r=0),
\end{equation}
where $\psi$ is the wave function, $r$ is the particle-particle
distance, $\hat{\psi}$ is the limit of the average value of the wave function on a
sphere centered at $r=0$ as its radius shrinks to zero, $\mu_{ij}$ is the
reduced mass of the two particles, and $q_i$ and $q_j$ are the charges
of the two particles.

Pack and Byers Brown \cite{PackByersBrown1966} extended the analysis to
show that the wave function could be expanded in terms of hydrogenic
solutions.
\begin{equation}
\label{PackByersBrown}
\psi=\sum_{lm}a_{lm}r^l Y_l^m[\theta,\phi]\left(1+\frac{q_i q_j \mu_{ij}}{l+1}r+O[r^2]\right),
\end{equation}
where $l\ge 0$, $|m|\le l$, $a_{lm}$
is an expansion coefficient, $\theta$ and $\phi$ are the usual
spherical angles giving the orientation of the two particles, and
$Y_l^m$ is the usual spherical harmonic.

These results describe the regularity required at the Coulomb
singularities.  There are three practical approaches to making sure
the solution has the appropriate behavior.
\begin{enumerate}
\item Behavioral

Assume that local solutions to the Schr\"{o}dinger equation that fail
to satisfy Eqs. (\ref{KatoCuspCond}) and (\ref{PackByersBrown}) are not analytic;
assume that the expansion (cardinal functions) employed in the
numerical treatment is incapable of representing this nonanalytic
behavior. Granted these assumptions,
all numerical solutions will automatically be regular at the
point in question \footnote{Because it is impossible to work where the
  potential diverges, the grid must be designed to exclude the point
  in question. The fact the grid does not contain the point is not a
  requirement of the behavioral approach.}.  According to Boyd
\cite{Boyd2000}, in many contexts this approach is sufficient. If
the solutions that do not satisfy Eqs. (\ref{KatoCuspCond}) and (\ref{PackByersBrown})
have only weakly singular behavior,
the convergence rate may be
slow.  

\item Regularity

Replace the Hamiltonian at the singular points with the Kato cusp
conditions without otherwise altering the domain. 
The cusp conditions
are
\begin{eqnarray}
\left.\frac{\partial\hat{\psi}}{\partial \phi}\right|_{\phi=0}&=&-Z\rho\psi(\phi=0)\\
\left.\frac{\partial\hat{\psi}}{\partial \phi}\right|_{\phi=\pi/2}&=&Z\rho\psi\left(\phi=\frac{\pi}{2}\right)\\
\left.\frac{\partial\hat{\psi}}{\partial \zeta}\right|_{\zeta=0}&=&\frac{\alpha\rho}{2}\psi(\zeta=0)
\end{eqnarray}
or
\begin{eqnarray}
\psi(\phi=0)=0\\
\psi\left(\phi=\frac{\pi}{2}\right)=0\\
\psi(\zeta=0)=0,
\end{eqnarray}
where these two sets are mixed and matched while preserving the appropriate
symmetry or antisymmetry. The choice depends on precisely which state one
wishes to calculate. 

\item Excision

Excise a small sphere around the singular points and
impose boundary conditions on its surface that yield the correct
behavior at the singularity as the sphere shrinks. 
From equation Eq. (\ref{PackByersBrown})
\begin{eqnarray}
\label{psi2Dexp}
\psi&=&\sum_l b_l\rho^l\phi^l P_l[C]\left(1-\frac{Z \rho\phi}{l+1}+O[\rho^2\phi^2]\right)\\
\psi&=&\sum_l c_l\rho^l\tilde{\phi}^lP_l[C]\left(1-\frac{Z\rho\tilde{\phi}}{l+1}
+O\left[\rho^2\tilde{\phi}^2\right]\right)\\
\label{zetaexp}
\psi&=&\sum_l d_l\rho^l\zeta^l P_l[B]\left(1+\frac{\alpha\rho\zeta\sqrt{2}}{2(l+1)}+O[\rho^2\zeta^2]\right),
\end{eqnarray}
where $\tilde{\phi}=\pi/2-\phi$, $P_l$ is a Legendre polynomial of order $l$, and $b_l$, $c_l$, and $d_l$ are unknown constants.
Define
\begin{eqnarray}
\label{Projection}
\xi_l &=& \int_{-1}^1 \psi P_l[C]\ud C\\
\chi_l &=& \int_{-1}^1 \psi P_l[B]\ud B,
\end{eqnarray}
and write the conditions as
\begin{eqnarray}
\label{ExcisionBC}
0&=&\phi \partd{\xi_l}{\phi}+\left(-l+\rho\phi\frac{Z}{l+1}\right)\xi_l\\
&=&-\tilde{\phi} \partd{\xi_l}{\phi}
+\left(-l+\rho\tilde{\phi}\frac{Z}{l+1}\right)\xi_l\\
\label{eeCusp}
&=&\zeta\partd{\chi_l}{\zeta}-
\left(l+\frac{\alpha\rho\zeta\sqrt{2}}{2(l+1)}\right)\chi_l.
\end{eqnarray}
These conditions become exact as the excised volume shrinks to a
point. For PS methods the volume should be reduced
  exponentially with increasing resolution.  The changes do not
  increase the computational cost, but may adversely affect the
  condition number of the matrix.

\end{enumerate}

The difficulty of implementation increases with number on the list. 

\subsection{\label{Fock}Three particle coalescence}

The potential is also singular when all three particles collide (i.e.,
when the hyperradius, $\rho$ goes to zero).  The behavior of the wave
function about this point is much less well understood than
two-particle coalescences and cannot be handled in the same way.
Instead of simply having a discontinuity in the wave function's first
derivative (the value of which is finite on both sides of the
singularity), the second derivative grows logarithmically near
$\rho=0$.  Bartlett \cite{Bartlett1937} was the first to show that a
simple Frobenius type expansion in powers of $\rho$ about $\rho=0$
fails at second order on account of the electron-electron interaction.
He suggested that logarithmic terms exist in the exact solution of
helium. Fock \cite{Fock1954,Fock1958} introduced an expansion of the
form
\begin{equation}
\psi=\sum_{n=0}^\infty\sum_{m=0}^{\lfloor n/2 \rfloor} e_{nm}\rho^n(\log \rho)^m,
\end{equation}
where $e_{nm}$ are two dimensional functions of the hyperangles
$\{\phi,C\}$ or $\{\zeta,B\}$ determined through the
recursive relationship
\begin{widetext}
\begin{equation}
\label{FockRecursive}
[n(n+4)+\triangle]e_{nm}=2Ve_{n-1,m}-2Ee_{n-2,m}
-2(n+2)(m+1)e_{n,m+1}-(m+1)(m+2)e_{n,m+2},
\end{equation}
\end{widetext}
and $\triangle$ is the two-dimensional Laplacian over the
hyperangles.  All $e_{nm}$ with $n \le 2$ are known analytically
plus a few additional terms with higher $n$
\cite{AbbottMaslen1987,GottschalkEtAl1987,GottschalkMaslen1987,McIsaacMaslen1987}.
Morgan proved that the series is convergent everywhere
\cite{Morgan1986}, and it has been shown that variational calculations
converge faster when a single power of a logarithm is included in the
basis \cite{FrankowskiPekeris1966} \footnote{Curiously, higher powers
  of the logarithm do not seem to improve the convergence rate of
  variational calculations \cite{Schwartz2006}.}.

Again, there are three basic strategies for a numerical scheme.
\begin{enumerate}
\item Behavioral

Do nothing special and rely on the regularity of the cardinal functions. This is an
imperfect approach since the exact wave function has unbounded second
derivatives as $\rho \to 0$.  If the basis set can only represent
regular behavior as discussed in the case of two-particle coalescence
it will not produce the exact solution.  However, since the volume
element scales as $\rho^5 d\rho$ such inexactness may have negligible
effect on observables calculated from the wave function.

\item Regularity

Impose a regularity-like condition at the singular point.  For the
ground state (and many other {\it S} states), the first-order solution \cite{Fock1954,Fock1958} 
to the Fock equations \ref{FockRecursive} is
\begin{eqnarray}
e_{00}&=&c\\
e_{10}&=&c \left\{-Z(\cos{\phi}+\sin{\phi})+\frac{\alpha}{2}\sigma[C,\phi]\right\},
\end{eqnarray}
where $c$ is a constant given by the normalization. These solutions imply either
\begin{equation}
\partial_\rho\psi|_{\rho=0}=\left\{-Z(\cos{\phi}+\sin{\phi})+\frac{\alpha}{2}\sigma[C,\phi]\right\}\psi(\rho=0),
\end{equation}
which is valid for the ground state, or
\begin{equation}
\psi(\rho=0)=0.
\end{equation}
Note that this regularity-like condition says nothing about the second
derivatives.  For the same reasons as above this method can never give the
exact solution at $\rho=0$.

\item Excision

Excise a small domain with $\rho<\rho_{\textrm{min}}$, where
$\rho_{\textrm{min}}$ is the cutoff.  A boundary condition can be
calculated on this inner surface by solving equation
\ref{FockRecursive} to relate the wave function and normal derivative on
the surface
\footnote{The boundary condition is not known in analytic form
(except at low order) but must be inferred by a numerical technique.
Solutions to equation
\ref{FockRecursive} can be found numerically with similar
techniques as employed here for the full three dimensional
problem.} 
.
As the resolution increases, more terms must be calculated
so that the truncation error in the Fock expansion equals the error
due to having finite resolution in the numerical calculation.
The basis set expansion in the bulk remains completely
regular. While exact, the method is complicated and will be pursued at
a later time.

\end{enumerate}

\subsection{Infinite separation}

The domain of the ideal problem extends to infinity. The bound-state
wave functions fall off exponentially. The outer boundary condition
must be approximated in the numerical method.  There are several
approaches.

\begin{enumerate}
\item Behavioral

Rely on the regularity of the cardinal functions to exclude
exponentially growing solutions as $\rho \to \infty$.
One must map the semi-infinite domain to a finite one to use Chebyshev
collocation points or work with semi-infinite functions like Laguerre
polynomials. 

\item Regularity

Again map the domain to a finite one but 
replace the Schr\"{o}dinger equation at $\rho=\infty$ by
\begin{equation}
\psi(\rho=\infty)=0.
\end{equation}
Note that if one includes the endpoints using the behavioral method, this 
method is effectively the same as the behavioral condition because the Schr\"{o}dinger
equation reduces to $E\psi = 0$ at $x=-1$ ($\rho=\infty$).

\item Excision

Excise the region with $\rho > \rho_{\textrm{max}}$ where
$\rho_{\textrm{max}}$ is the cutoff and impose a suitable
boundary condition. As in the three-particle
coalescence, one may develop a more and more accurate
representation at fixed $\rho_{\textrm{max}}$ and/or an approximate
condition at increasing $\rho_{\textrm{max}}$. It is easiest to
set
\begin{equation}
\psi(\rho=\rho_{\textrm{max}})=0
\end{equation}
or
\begin{equation}
\left.\frac{\partial\psi}{\partial\rho}\right|_{\rho=\rho_\textrm{max}}=0
\end{equation}
and vary $\rho_{\textrm{max}}$, which is what is done in this article when using this method.  
In a PS numerical scheme one should
vary $\rho_{\textrm{max}} \propto n^{1/2}$ for large $n$ where $n$ is the radial
resolution (see Appendix \ref{Truncation}).

\end{enumerate}

\subsection{Collinearity ($B$ or $C=\pm 1$)}

The coefficients multiplying the second derivatives with respect to
$B$ and $C$ at $B$,$C=\pm 1$ go to zero.  In ordinary differential
equations this allows irregular solutions that behave as linear
combinations of Legendre functions of the second kind.  Regularity of
the cardinal functions excludes such solutions. Since the
Schr\"{o}dinger equation at these points contains no infinities it
does not matter if the grid includes these points.  The partial
differential equation is parabolic along this boundary. So no 
boundary conditions or regularity conditions need to be 
given.

\section{The Pseudospectral Method}

Boyd \cite{Boyd2000}, Fornberg \cite{Fornberg1996}, Pfeiffer {\it et
al} \cite{PfeifferEtAl2003}, and the third edition of {\it Numerical
Recipes} \cite{NumericalRecipes} cover the PS method in
detail. A brief review of some aspects pertinent to our work follows.

The main advantage of this method is that it provides exponentially
fast convergence for smooth solutions.  Unlike finite difference and
finite element algorithms, all derivatives are calculated to higher
and higher order with increasing resolution.

It is also noteworthy that the grid points are clustered more closely near
the boundary of a domain than in its center. 
With this arrangement the representation of a function and its
derivative is more uniformly accurate across the whole domain than is
possible using an equal number of equidistant points.  Finite
difference and finite element methods typically use an equal-spaced
grid and the derivatives are less accurate at the edge than at the
center.

Let $n_d$ be the number of coordinate dimensions and $N_i$ the resolution in the
$i$th dimension.  The differential equation is enforced at
$n_t=\prod_{i=1}^{n_d} N_i$ collocation or grid points chosen to be the
roots or extrema of a Jacobi polynomial of order $N_i$ in each dimension.
Boyd's recommendation that one use Chebyshev polynomials to generate
the grid points in lieu of special circumstances is followed here
\cite{Boyd2000}.

The derivatives in the $i^\textrm{th}$ direction are
calculated to $N_i^\textrm{th}$ order in terms of the function values
at the collocation points.  To illustrate this it is useful to define
the cardinal functions:
\begin{equation}
C_j^N[x]=\prod_{\substack{i=1\\i\ne j}}^N \frac{x-x^i}{x^j-x^i},
\end{equation}
where the $x^i$'s are the collocation points and $i$ and $j$ are
superscripts not exponents.  These functions have the property that
\begin{equation}
C_j^N[x^i]=\delta_j^i.
\end{equation}
The PS representation of a function at an arbitrary $n_d$-dimensional 
position $(x_1,\ldots,x_{n_d})$ is expanded as
\begin{equation}
\label{PSexpansion}
\psi\approx\psi_{N_1,\ldots,N_{n_d}}= \psi^{j_1,\ldots,j_{n_d}}F_{j_1,\ldots,j_{n_d}}^{N_1,\ldots,N_{n_d}}
[x_1,\ldots,x_{n_d}],
\end{equation}
(using the Einstein summation convention) in terms of its
grid values and the cardinal functions
\begin{eqnarray}
\psi^{j_1,\ldots,j_{n_d}}&=&\psi_{N_1,\ldots,N_{n_d}}[x_1^{j_1},\ldots,x_{n_d}^{j_{n_d}}]\\
F_{j_1,\ldots,j_{n_d}}^{N_1,\ldots,N_{n_d}}&=&\prod_{i=1}^{n_d} C_{j_i}^{N_i}[x_i] .
\end{eqnarray}
Such an expansion is equivalent (up to an exponentially small error for smooth functions) to a spectral one,
\begin{equation}
\psi=s^{j_1,\ldots,j_{n_d}}G_{j_1,\ldots,j_{n_d}}
[x_1,\ldots,x_{n_d}],
\end{equation}
where
\begin{eqnarray}
G_{\bd{j}}&=&\prod_{i=1}^{n_d} u_{j_i}[x_i]\\
s^{j_1,\ldots,j_{n_d}}&=&\int \psi G_{\bd{j}} \prod_{i=1}^{n_d} w[x_i] dx_i\\
&\approx&\psi^{i_1,\ldots,i_{n_d}}G_{\bd{j}}[x_1^{i_1},\ldots,x_{n_d}^{i_{n_d}}]\prod_{k=1}^{n_d} v_{i_k},
\end{eqnarray}
$\bd{j}$ means $j_1,\ldots,j_{n_d}$,
$u_{j_i}$ are orthonormal polynomials chosen here to be Chebyshev polynomials, 
$w$ is the weight function over which they are orthonormal,
and $v_{i_k}$ are the corresponding quadrature weights.
Note, the collocation points are also quadrature points which allow exponential 
convergence of the quadrature. 
Derivatives of
$\psi$ are approximated by differentiating Eq. (\ref{PSexpansion}). To this end, it is 
useful to introduce differentiation matrices,
\begin{equation}
(D_N)_j^i=\partial_xC_j^N[x]|_{x=x^i}.
\end{equation}

This method is readily applied to linear eigenvalue problems
as arise from the Schr\"{o}dinger equation
\footnote{Linearity is not required for the PS method.}
\begin{equation}
({\cal{H}}-E)\psi=0,
\end{equation}
where $\cal{H}$ is the Hamiltonian 
operator, $\psi$ is the wave function, and $E$ is the energy eigenvalue. 
Discretizing on the grid gives the tensor equation
\begin{equation}
H_{j_1,j_2,\ldots,j_{n_d}}^{i_1,i_2,\ldots,i_{n_d}}\psi^{j_1,j_2,\ldots,j_{n_d}}=E\psi^{i_1,i_2,\ldots,i_{n_d}},
\end{equation}
where 
\begin{equation}
H_{j_1,j_2,\ldots,j_{n_d}}^{i_1,i_2,\ldots,i_{n_d}}={\cal{H}}F_{j_1,j_2,\ldots,j_{n_d}}
[x_1^{i_1},x_2^{i_2},\ldots,x_{n_d}^{i_{n_d}}] .
\end{equation}
Unlike finite difference methods, the tensor
$H_{j_1,j_2,\ldots,j_{n_d}}^{i_1,i_2,\ldots,i_{n_d}}$ is dense. Write this as
$H_{l(j_1,j_2,\ldots, j_{n_d})}^{k(i_1,i_2,\ldots,i_{n_d})}$ or for short
$H_l^k$, where $k$ and $l$ are one-to-one functions mapping the set of
$n_d$ indices to the lowest $n_t$ positive integers. This recasts the
tensor as a large matrix so that standard matrix methods can be
employed.

One way to carry out the mapping employs the Kronecker product as
follows. If ${\cal{H}}$ is given by
\begin{equation}
{\cal{H}}=\sum_{i_1\ldots,i_{n_d}} f_{i_1,\ldots,i_{n_d}}[x_1,\ldots,x_{n_d}](\partial_{x_1})^{i_1}\cdots(\partial_{x_{n_d}})^{i_{n_d}},
\end{equation}
where $f_{i_1,\ldots,i_{n_d}}$ is a function coefficient,
then the matrix, $H$ is given by
\begin{equation}
H^k_l=\sum_{i_1,\ldots,i_{n_d}}f_{i_1,\ldots,i_{n_d}}[\bd{x}^k]\left(\bigotimes_{j=1}^{n_d} (D_{N_j})^{i_j}\right)^k_l,
\end{equation}
where $\bd{x}^k$ is the $n_d$-dimensional vector of coordinates with indices that map to $k$,
$N_j$ is the number of grid points in the $j^\textrm{th}$ direction, $i_j$ is an exponent, and
$D_{N_j}$ is the differential matrix based on $N_j$ points.

\section{Pseudospectral Convergence for NonSmooth Functions}

To make appropriate design algorithmic choices it is important
to investigate how the PS method handles
nonsmooth behavior of solutions. This section explores the
convergence of truncated cardinal function expansions to
cusps and logarithmic terms and then employs a toy model
that illustrates how the triple coalescence is expected to
influence the numerical results.

\subsection{Kato cusps}

\begin{figure}
\includegraphics[width=\linewidth]{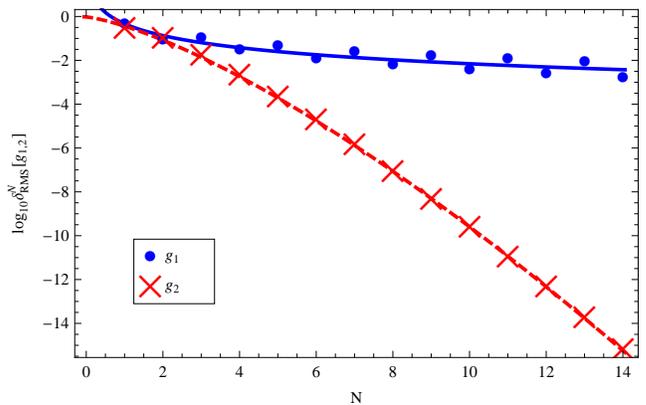}
\caption{\label{CuspErr} (Color online). The logarithm base 10 of 
$\delta_{\textrm{RMS}}^N[g_1]$ (blue circles) and 
$\delta_{\textrm{RMS}}^N[g_2]$ (red crosses) with solid
blue and dashed red fits, respectively. See Appendix \ref{FitsSection} for
fitting functions.}
\end{figure}

Consider the ground state of the hydrogen atom
with wave function
\begin{equation}
\psi=e^{-r}=e^{-\sqrt{x^2+y^2+z^2}}.
\end{equation}
In Cartesian coordinates, there is
a discontinuity in the first derivative at the origin,
\begin{equation}
\lim_{x\rightarrow 0^+}\left.\frac{\partial\psi}{\partial x}\right|_{y=z=0}
\ne\lim_{x\rightarrow 0^-}\left.\frac{\partial\psi}{\partial x}\right|_{y=z=0}.
\end{equation}
In spherical coordinates no discontinuity exists for $r \ge 0$.
All the derivatives at $r=0$ are well defined 
and a PS code has no problem exponentially
converging toward the correct answer.
The essence of this observation can be seen by
considering the one-dimensional exponential functions
\begin{eqnarray}
g_1[x]&=&e^{-|x|}\\
g_2[x]&=&e^{-(x+1)},
\end{eqnarray}
on the domain $-1\le x\le 1$ with weight $x^2$ (analogous to the
three dimensional hydrogen atom).
As a measure of error between the 
function $f$ and its cardinal expansion
truncated at order $n$ define
\begin{equation}
\label{eqn-definermserror}
\delta_{\textrm{RMS}}^N[f]=\sqrt{\int_{-1}^1 x^2 \left(f[x]-\sum_{i=1}^N C_i^N[x] f[x^i]\right)^2 \ud x} .
\end{equation}
Figure \ref{CuspErr} compares $\delta_{\textrm{RMS}}^n[g_1]$ to
$\delta_{\textrm{RMS}}^n[g_2]$ as a function of $n$. Evidently the
cusp is poorly represented compared to the smooth function at a given $n$.
The PS representation of the cusp converges algebraically while the representation
of the smooth function converges supergeometrically (see Appendix \ref{FitsSection} for fits).

The basic strategy in more complicated problems is to adopt a
coordinate system with a radial-like coordinate at each cusp.  For
two-electron atoms no global coordinate system exists with the desired
property at each of the three separate two-particle coalescences. This
article uses three individual but overlapping domains to guarantee
appropriate treatment near each coalescence point.

\subsection{\label{OneDLogErr}Logarithmic terms}
\begin{figure}
\includegraphics[width=\linewidth]{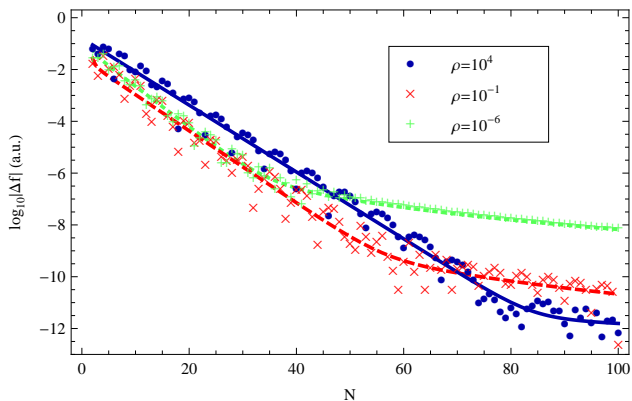}
\caption{\label{LogErr} (Color online). The logarithm base 10 of the error in $f[x]$ using cardinal functions at 
$\rho=10^{-6}$ (green pluses), $\rho=10^{-1}$ (red crosses), and $\rho=10^4$ (blue circles)
with solid blue, dashed red, and dotted green fits, respectively. See Appendix \ref{FitsSection}
for fitting functions and method.
}
\end{figure}

Consider the one-dimensional function
\begin{equation}
\label{OneDFunc}
f[x]=\left(1+\frac{1}{2}\alpha\rho[x]^2\log[\rho[x]]\right)e^{-\rho[x]},
\end{equation} 
where
\begin{equation}
\rho[x]=\frac{1-x}{1+x}.
\end{equation}
Here $\rho\in [0,\infty)$, $x\in [-1,1]$ and $f$ are 
analogous to the hyperspherical radius, its algebraic transformation and 
the heliumlike wavefunction $\psi$, respectively. As in the full three-dimensional 
problem, the presence or absence of the logarithmic terms is controlled by $\alpha$, which
can be set to $0$ or $1$.

There are two types of errors considered here: interpolation error and operator error. 
These are different sorts of 
error, but qualitative features (e.g., exponential or algebraic convergence) are expected to be the same.

\subsubsection{Interpolation error}

The pointwise error  
between $f$ and its truncated expansion is
\begin{equation}
\Delta f=f[x]-\sum_{i=1}^N C_i^N[x] f[x^i],
\end{equation}
where $x^i$ refers to the $i$th grid point.

Figure \ref{LogErr} shows that the behavior of $\Delta f$ at three different 
values of $\rho$. For each value the apparent rate of convergence starts out
exponential before becoming algebraic at large $N$. 
The algebraic convergence known with the highest accuracy in Fig. \ref{LogErr} is for $\rho=10^{-6}$,
which asymptotically goes as $1/N^{3.82\pm 0.09}$ (see Appendix \ref{FitsSection}). 
This algebraic behavior is expected when trying to represent a nonanalytic function ($\log[\rho]$)
with an analytic basis. Such behavior disappears if $\alpha$ is set to zero.

The onset of algebraic convergence varies from $N\approx 40$ to
$N \approx 80$ as $\rho$, moving away from the singularity,
increases by 10 orders of magnitude.
The error at the transition is  $< 10^{-6}$.
Typically, energy errors vary as the square of wave function errors
and
would already be very small compared to relativistic corrections. This calculation shows that 
it is possible to get precise values with apparent exponential convergence
before reaching the asymptotic algebraic regime. As a practical matter,
one may never reach the latter limit.

\subsubsection{Operator error}

In order to estimate the error in the eigenvalue, it would help to have 
a one-dimensional toy eigenvalue problem with an eigenfunction similar to 
the function in Eq. (\ref{OneDFunc}).
It is impossible to construct a one-dimensional eigenvalue problem with
solutions that have logarithmic singularities without explicitly introducing
such singularities into the differential operator. So here a more limited test
problem is used. Instead of solving for an eigenvalue, the error in the 
operator is measured. This would contribute to the eigenvalue error along
with the error in the wave function.

Let $\Delta {\cal{H}}$ be the difference between the true Hamiltonian
and the Hamiltonian constructed from PS differentiation matrices. The
associated energy error is $\langle\psi |\Delta
{\cal{H}}|\psi\rangle$.  The aim is to construct
an analog of the integrand of the energy error and use it to assess
pointwise and integral errors.

Construct a differential operator ${\cal D}$ similar to
the full Hamiltonian ${\cal H}$ [see Eq. (\ref{Hamiltonian})]
but in terms of the coordinate x,
\begin{equation}
{\cal{D}}=p_2[x]\partial_{xx}+p_1[x]\partial_x+p_0[x],
\end{equation}
where
\begin{eqnarray}
p_2[x]&=&-\frac{(1+x)^4}{8}\\
p_1[x]&=&\frac{(1+x)^3}{4}\left(\frac{4+x}{1-x}\right)\\
p_0[x]&=&-\frac{1}{\rho [x]}.
\end{eqnarray}
Note, the first two terms are identical to the operator ${\cal{T}}_\rho$ and the last term
is a Coulomb potential.

The corresponding matrix operator is
\begin{equation}
(d_N)_j^i=p_2[x^i](D_{N})_k^i(D_{N})_j^k+p_1[x^i](D_{N})_j^i+p_0[x^i]\delta_j^i,
\end{equation}
where Einstein's summation convention is used and 
$\delta_j^i$ is the Kronecker delta function.
The pointwise error on the grid and its maximum are
\begin{eqnarray}
(\Delta H_N)^i &=& w[x^i]\left(({\cal{D}}f)[x^i]-\sum_j(d_N)_j^i f[x^j]\right)\\
\Delta H_N^{\textrm{max}}&=&\max_i |(\Delta H_N)^i|,
\end{eqnarray}
where 
\begin{equation}
w[x]=\frac{(1-x)^5}{(1+x)^7}f[x]
\end{equation}
The factor  $(1-x)^5/(1+x)^7$ comes from the Jacobian. 

\begin{figure}
\includegraphics[width=\linewidth]{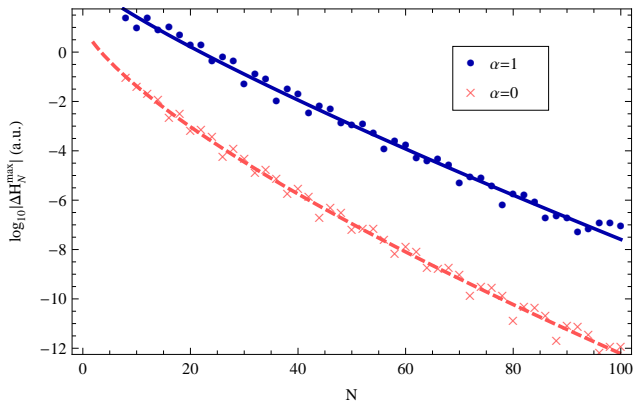}
\caption{\label{DHMax} (Color online). The logarithm base 10 of the maximum error of a pseudospectral matrix. The dark blue circles are for $\alpha=1$
and the light red crosses for $\alpha=0$ with solid blue and dashed red fits, respectively. See Appendix \ref{FitsSection} for fitting functions and method.}
\end{figure}
\begin{figure}
\includegraphics[width=\linewidth]{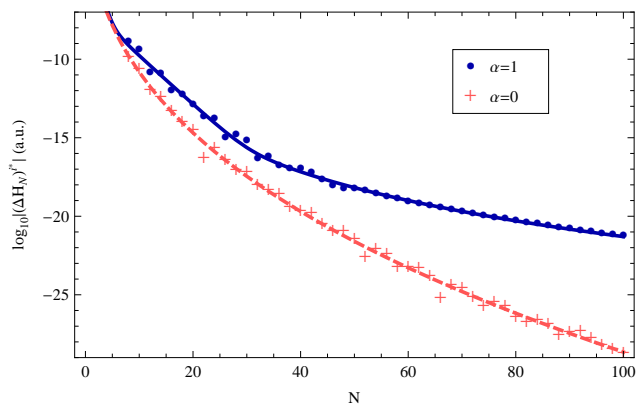}
\caption{\label{DHIn} (Color online). The logarithm base 10 of the error of a pseudospectral matrix near the singularity. 
The dark blue circles are for $\alpha=1$
and the light red pluses for $\alpha=0$ with solid blue and dashed red fits, respectively. See Appendix \ref{FitsSection} for fitting functions and method.}
\end{figure}

Figure \ref{DHMax} shows $\Delta H_N^{\textrm{max}}$, the maximum
error anywhere on the grid, as a function of $n$ for $\alpha=0$ and
$\alpha=1$.  The decrease appears to be exponential, not unanticipated when $\alpha=0$
but perhaps a surprise for $\alpha=1$.  The slopes of the two curves are
roughly the same and the offset is due to the variation of the
magnitude of $f$ with $\alpha$. An explanation is immediately
suggested by Fig. \ref{DHIn} which shows the error at the grid point
closest to the singularity $(\Delta H_N)^{i^*}$ (here $i^*$ refers to
that point).  The data for $\alpha=1$ is well fit by an algebraic rate
of convergence ($1/N^{10.36\pm 0.08}$) at large $N$ while $\alpha=0$ has an
approximately exponential fall-off (the convergence is subgeometric because
the calculation is done on a semi-infinite domain). The log term {\it
  does} spoil the method's exponential convergence. Assuming that the
effect is greatest at $i=i^*$ , the maximum error is
dominated by the log term when $N$ is greater than about $200$ and the error
is {\it very} small. This is exponential convergence ``for all
practical purposes.'' 
%
%
%

\subsection{Conclusion}

For the interpolation and operator errors, the logarithmic term does 
not slow convergence unless one is at high resolution or interested 
in small values of $\rho$. For those cases, one would need to apply 
the excision method about the triple coalescence point in order to 
retain exponential convergence. For most applications the level of precision
needed is obtained before the algebraic behavior becomes apparent.

\section{Numerical Domains and Collocation Points}

Because the two electrons are identical particles the full wave function
is antisymmetric. In the ground state, the spins are antisymmetric
and the spatial part is symmetric. The spatial domain may be taken
as $\phi<\pi/4$ and $B>0$. The complete numerical domain is
\begin{equation}
\begin{array}{lccc}
D_0:&-1\le x\le 1 & 0\le\phi\le\frac{\pi}{4} & -1\le C\le 1\\
&&\textrm{or}&\\
D_0:&-1\le x\le 1 & 0\le\zeta\le\frac{\pi}{2} & 0\le B\le 1.
\end{array}
\end{equation}
A single domain does not allow the proper treatment of the two-particle
coalescences. Therefore, introduce three subdomains to cover $D_0$
using two different sets of variables:
\begin{figure}
\includegraphics[width=\linewidth]{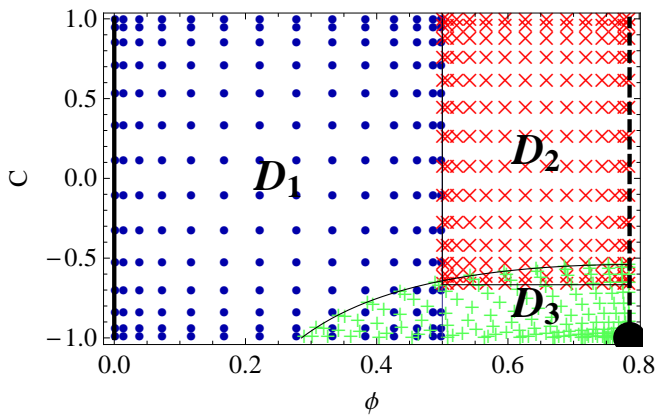}%
\caption{\label{GridPoints} (Color online).
This is the arrangement of grid points of the three domains at a
constant value of $\rho$ in $\phi$ and $C$ coordinates. Note that the
point density becomes larger at the boundary of each subdomain and
that no grid points sit on the Coulomb singularities. The blue circles,
red crosses, and green pluses belong to domains $D_1$, $D_2$, and $D_3$,
respectively. $D_1$ and $D_2$ are rectangular domains, while $D_3$ has the 
curved boundary in $\phi$, $C$ coordinates but is rectangular in $\zeta$, $B$ coordinates. 
The electron-proton singularity occurs on the left side
(solid line at $\phi=0$). The entire line corresponds to one physical
point. The electron-electron singularity occurs at the lower right-hand
corner (solid disk at $\phi=\pi/4, C=-1$). A line of symmetry falls on
the right side (dashed line at $\phi=\pi/4$ where $r_1=r_2$).}
\end{figure}
\begin{equation}
\begin{array}{lccc}
D_1:&-1\le x\le 1 & 0\le\phi\le\frac{1}{2} & -1\le C\le 1\\
D_2:&-1\le x\le 1 & \frac{1}{2}\le\phi\le\frac{\pi}{4} & -\frac{2}{3}\le C\le1\\
D_3:&-1\le x\le 1 & 0\le\zeta\le\frac{1}{2} & 0\le B\le 1.
\end{array}
\end{equation}
For calculations done on a finite domain, the condition $-1\le x\le 1$ is replaced
by $0\le \rho\le\rho_\textrm{max}$. Cross sections of these domains at fixed $\rho$
are shown in Fig. \ref{GridPoints}.  An
electron-proton singularity lies in $D_1$, while the
electron-electron singularity lies in $D_3$.  The radial-like
coordinates in $D_1$ ($\phi$) and $D_3$ ($\zeta$) accommodate the
cusps just like the usual radial coordinate does in the hydrogen atom.
$D_2$ fills in the remaining volume. All three domains have boundaries
that touch the triple coalescence point (not pictured).
%

Consideration of
the electron-electron singularity shows why the single domain $D_0$ is inadequate. Byers Brown and White
\cite{ByersBrownWhite1967}
showed that the wavefunction can be expanded in powers of $r_{12}$ about $r_{12}=0$. 
Using such a coordinate accurately treats the cusp away from the triple coalescence point.
The expansion in powers of $\zeta$ is very similar [see Eq. (\ref{zetaexp})] or
equivalently powers of $\sqrt{2}\sin\zeta=\sqrt{1+C\sin 2\phi}$ \footnote{
The radius of convergence of the Byers Brown and White expansion is unknown to the authors but is clearly invalid at $\rho=0$,
the location of the triple 
coalescence point. Here only the effect of the double coalescence point 
is being considered.}.
Re-expanding in $\phi$ and $C$ coordinates gives
derivatives of $\sqrt{1+C\sin 2\phi}$ with respect to $C$ and $\phi$,
terms that
are either infinite or undefined at $\phi=\pi/4$ and $C=-1$. 
This is why the PS method fails to converge
rapidly using $D_0$ alone.


Within each domain, grid points are set as follows.
Let the $i$th dimension extend from $x_{i,\textrm{min}}$
to $x_{i,\textrm{max}}$ and have $N_i$ collocation
points.  These points are the roots or antinodes plus endpoints of the
$N_i^\textrm{th}$ order Chebyshev polynomial 
\footnote{When the excision method is used near the two-particle
coalescence points, the nodes of Legendre polynomials are used in the $B$
and $C$ directions. This choice makes it easier to apply Eq. 
(\ref{ExcisionBC}) with a simple quadrature.}
stretched to fit length
$\Delta x_i = x_{i,\textrm{max}}-x_{i,\textrm{min}}$.  The
$j$th point (for $j=1,2,\ldots,N_i$) of dimension $i$ is
\begin{equation}
x_i^j=\frac{\Delta x_i}{2}(y_i^j+1),
\end{equation}
where
\begin{equation}
y_i^j=\cos\left[\frac{(N_i-j+\lambda)\pi}{N_i}\right]
\end{equation}
and $\lambda$ is $0$ or $1/2$ for nodes
or antinodes plus endpoints, respectively 
\footnote{It is also possible to use the so called Chebyshev-Randau
points, which include one endpoint on one side of the domain but not
the other.}. 
In this article, 
nodes are generally used except when explicit boundary conditions
are needed at both endpoints,
$x_{i,\textrm{min}}$ and $x_{i,\textrm{max}}$. 

Potentially each dimension and domain could have its own $N_i$ but in
this paper the $x$ direction is set to be twice as large as the other
two dimensions and all are varied in lockstep.  That is,
$\{N_x,N_C,N_\phi\}=\{2n,n,n\}$ in domains $D_1$ and $D_2$ and
$\{N_x,N_B,N_\zeta\}=\{2n,n,n\}$ in domain $D_3$.  The total number
of grid points is $n_{t}=3\times(2n\times n\times n)=6n^3$ points.
Twice as many points were used in the $x$ dimension, an arbitrary
choice but one motivated by the semi-infinite range of the
hyperspherical coordinate and by the wave function's logarithmic
dependence on the hyperspherical radius near the triple coalescence
point.

%

\section{Boundary Conditions}

\subsection{Internal boundary conditions}

It is necessary to ensure continuity of
the wave function and its normal derivative at internal boundaries.
There are two ways in which the subdomains can touch:
they can overlap or they can barely touch.
For clarity, consider a one-dimensional problem with two domains.
Let the first domain be domain $1$ and the second be 
domain $2$ with extrema 
$x_{1,\textrm{min}}<x_{2,\textrm{min}}\le x_{1,\textrm{max}}<x_{2,\textrm{max}}$,
where the $1$ and $2$ now refer to domain number.
The first case corresponds to $x_{2,\textrm{min}}<x_{1,\textrm{max}}$ 
and the second to $x_{2,\textrm{min}}=x_{1,\textrm{max}}\equiv x_*$.
For both cases, exactly two conditions are need to make the wave function
and its derivative continuous. The simplest choice for the first case is
\begin{eqnarray}
\psi_1[x_{1,\textrm{max}}]&=&\psi_2[x_{1,\textrm{max}}]\\
\psi_1[x_{2,\textrm{min}}]&=&\psi_2[x_{2,\textrm{min}}],
\end{eqnarray}
and for the second case is
\begin{eqnarray}
\psi_1[x_{*}]&=&\psi_2[x_{*}]\\
\frac{d}{dx}\psi_1[x_{*}]&=&\frac{d}{dx}\psi_2[x_{*}].
\end{eqnarray}

\begin{figure}
\includegraphics[width=\linewidth]{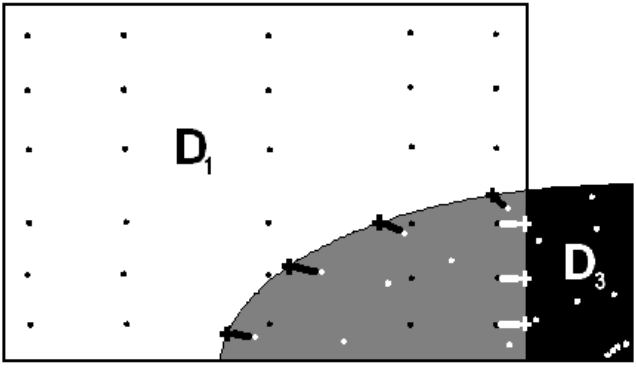}%
\caption{\label{Overlap} The intersection (gray) of domains $D_1$ (white) 
with black grid points and $D_3$ (black) with white grid points. The boundary
points are depicted as black and white crosses and are connected via 
black and white lines to the grid points that they replace.}
\end{figure}

\begin{figure}
\includegraphics[width=\linewidth]{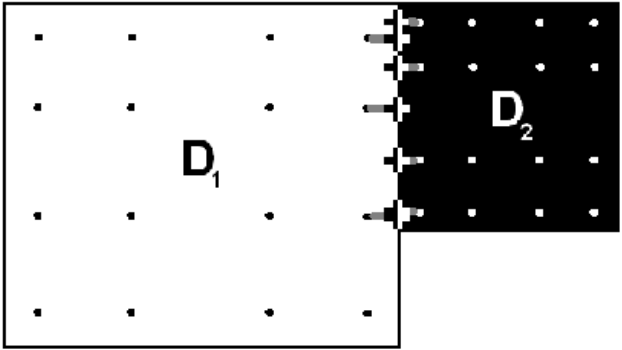}%
\caption{\label{Touch} Barely touching domains $D_1$ (white) 
with black grid points and $D_2$ (black) with white grid points. The boundary
points are depicted as black and white crosses and are connected via 
gray lines to the grid points that they replace.}
\end{figure}

For multidimensional grids, the situation is analogous.
The conditions are applied 
on surfaces and the derivatives are normal derivatives at the surface.
On a discrete grid, a finite number of conditions
are given which, in the limit of an infinitely fine mesh,
would cover the entire surface. There is a great deal of freedom
in the selection of the points but
in this article the edge of a domain has one constant
coordinate so there is a natural choice. Conditions are
imposed at the points of the finite mesh formed by varying
all the other coordinates (in general, these are not collocation
points). In other words, the matching points lie at
the intersection of the coordinate lines normal to the surface
with the surface itself.
The positions of the crosses in Figs. \ref{Overlap}
and \ref{Touch} illustrate where the matching occurs
when the domains overlap and when they just touch.

For touching domains, the black and white crosses in Fig.
\ref{Touch} are used. Note that four (three) crosses are defined by
the coordinate lines in $D_2$ ($D_1$).  At the set of four crosses,
function values are equated, and at the set of three crosses,
normal derivatives are equated. In general, function values (derivatives)
are equated at points stemming from the subdomain with a greater (lesser)
density of points along the boundary.

For overlapping domains, the function values are
equated at the black and white crosses in Fig. \ref{Overlap} which
lie on two separate surfaces. The points are selected in
a manner similar to that of touching domains, i.e. in terms of the
intersection of the coordinate lines in $D_1$ and $D_3$ with the
surface.

In all cases, values and derivatives at all points are calculated
using Eq. (\ref{PSexpansion}) and are ultimately linear
combinations of the grid point values.


\subsection{Symmetry and regularity conditions}

For this problem there are two types of boundary conditions on 
the boundary of the numerical domain: the symmetry
condition from electron exchange and the regularity conditions
imposed near singular points.

The
symmetry condition is related to the total spin of the two electrons,
$S$. If $S=0$ ($S=1$) the wave function is symmetric (antisymmetric)
about $\phi=\pi/4$ or $B=0$ and the normal derivative (value) of the wave
function is equal to zero. This condition is enforced
at all the points on the boundary that have the same $\rho$ and $C$
coordinates as grid points in $D_2$ or the same $\rho$ and $\zeta$
coordinates in $D_3$. This gives $2\times 2n\times n =4n^2$ boundary
conditions. 

Regularity conditions are imposed as boundary conditions at four
two-dimensional surfaces: $\rho=0$, $\rho=\infty$, $\phi=0$, and
$\zeta=0$. These are similar in form to the symmetry condition except
involve linear combinations of derivative and value.  Depending on which type
of conditions are given at singular points (behavioral versus regularity or excision)
are used, there are $0$
to $10 n^2$ conditions. These conditions replace an equal number of
equations. 
The particular equation replaced is the one that stems from 
enforcement of the discretized Schr\"{o}dinger equation at 
the collocation point nearest to the boundary at which the 
condition applies.

The most complicated type of boundary condition arises when 
a region is excised about a two-particle coalescence.
First, one must project out terms proportional 
to each Legendre polynomial by performing an integral over
$C$ or $B$. This can be done by quadrature over the grid points
in those dimensions. For example, Eq. (\ref{Projection})
turns into
\begin{equation}
\xi_l[\rho^i,\phi^j]=\sum_k w_k P_l[C^k]\psi[\rho^i,\phi^j,C^k],
\end{equation}
where $w_k$ are the quadrature weights. Then Eq. (\ref{ExcisionBC})
becomes for $j=1$ (the excision boundary)
\begin{equation}
0=\left(\phi^j (D_{N_\phi})_k^j
+\left(-l+\frac{\rho^i\phi^j Z}{l+1}\right)\delta_k^j\right)
\xi_l[\rho^i,\phi^k],
\end{equation}
which is $N_\rho N_C$ conditions
($0\le l \le N_C-1$ and $1\le i \le N_\rho$).

\subsection{Incorporating boundary conditions into the matrix problem}

All of the above boundary conditions are expressed as a linear combination of the 
function values at the grid points equal zero. In matrix form 
\begin{equation}
\label{BCMatEq}
n_b \{\underbrace{\left(B_1 B_2\right)}_{n_b+n_i}
\left(\begin{array}{c} \psi_1 \\ \psi_2 \end{array}\right)
\begin{array}{c} \}n_b \\ \}n_i \end{array} = 0,
\end{equation}
where $\psi_1$ ($\psi_2$) is a vector of the $n_b$ ($n_i$) wave function values
at all the boundary (interior) points, the boundary condition 
matrix has been broken into an $n_b$ by $n_b$ matrix $B_1$ and an
$n_b$ by $n_i$ matrix $B_2$, and $n_b+n_i=n_t$. 
For the case where an endpoint is not a collocation point,
the grid point nearest to the boundary, at which 
an explicit boundary condition is given, is considered as a boundary point. 
All the points near where behavioral boundary 
conditions are given are not included in this definition. These points 
are the ones that give rise to the first $n_b$ rows in Eqs. 
(\ref{BCMatEq}) and (\ref{HamMatEq}). Note, that this ordering was chosen
for clarity in this section.

There is also the Hamiltonian matrix equation
\begin{equation}
\label{HamMatEq}
\begin{array}{c} n_b \{ \\ n_i \{ \end{array}
\underbrace{\left(\begin{array}{c} H_{11}-E \\ H_{21} \end{array}
\begin{array}{c} H_{12} \\ H_{22}-E \end{array}\right)}_{n_b+n_i}
\left(\begin{array}{c} \psi_1 \\ \psi_2 \end{array} \right) = 0,
\end{equation}
where the Hamiltonian matrix has also been divided into four matrices:
$H_{11}$, $H_{12}$, $H_{21}$ and $H_{22}$.

So there are $n_t+n_b$ equations and $n_t$ unknowns ($\psi_1$ and $\psi_2$) 
as well as the eigenvalue. One could approximately solve these equations with 
singular value decomposition \cite{NumericalRecipes}, but it is much faster to 
simply discard the first $n_b$ rows of the Hamiltonian matrix and incorporate the boundary 
conditions into the remaining eigenvalue problem:
\begin{equation}
\label{EigenEquation}
(H_{22}-H_{21}B_1^{-1}B_2-E)\psi_2=0,
\end{equation}
where $B_1$ has an inverse because all of its rows are linearly independent
(otherwise more than one boundary condition would have been specified for a given
boundary point). Calculating the inverse is not too computationally 
expensive because $n_b\ll n_t$.
Then solve for $\psi_1$ afterwards with
\begin{equation}
\psi_1=-B_1^{-1}B_2\psi_2.
\end{equation}

\section{Matrix Methods}

In one dimension the Hamiltonian matrix for the PS method
is dense but in three dimensions with three domains the number of
nonzero elements scales as $24n^4$ out of a possible $36 n^6$. The
boundary condition matrix is also sparse with $8n^4$ non-zero elements
out of $48n^5$ (for the simplest case where behavioral conditions are 
used whenever possible).  Therefore, any
attempt to solve these equations should take advantage of these memory
savings.

Equation (\ref{EigenEquation}) is solved by the method of inverse
iteration \cite{NumericalRecipes} after shifting the eigenvalue with
an approximately known value. In cases where the exact eigenvalue is
not known {\it a priori}, one solves the full eigenvalue problem for a low
resolution case first and then at each successive iteration shifts the
eigenvalue using the result of the previous iteration.  Because the
Hamiltonian matrix is not symmetric, a complex eigenvalue may
occur. There is no theoretical reason prohibiting the numerical
eigenvalue from containing an imaginary part at finite resolution but,
in fact, none were generated for $n>5$. Of course, the imaginary
part contributes to the error which must converge to zero.

The above solution method can yield highly oscillatory wave functions which appear to diverge on the boundaries of the computational domain. These nonphysical wave functions do not satisfy the first $n_b$ rows of Eq. (\ref{HamMatEq}) and arise as an artifact of solving a subset of equations of the overdetermined system. They are easily identified and rejected and in no way affect the true solution.

The entire calculation for $n=14$ took only about 20 minutes on a
6-GHz machine. Memory needed to solve the linear equations was the
limiting factor because inverting the equation has requirements
scaling as $n^6$. 
The {\it g}eneralized {\it m}inimal {\it res}idual (GMRES) algorithm \cite{BarrettEtAl1994}
might reduce the memory requirements 
of the solution of the linear equations that arise in the inverse iteration.

For simplicity of coding, the above calculations were done using 
{\it Mathematica} \cite{Mathematica2007}. Care was taken to use predefined functions whenever
such choice was more efficient.

\section{Results}
\label{results}

This article is an exploration of the PS method as applied to
heliumlike systems, not an attempt to improve the energy
eigenvalues for bound states. That has already been done to a higher
precision than will ever be needed 
\cite{BakerEtAl1990,Drake1999,Korobov2000,Frolov2000,Korobov2002,DrakeEtAl2002,Frolov2006,
  Frolov2007,NakashimaNakatsuji2007,NakashimaNakatsuji2008,KurokawaEtAl2008}.
The focus here is on showing the PS method works in a new application
and assessing its convergence properties.

Table \ref{DifferentRuns} gives a list of runs used in this section to
discuss the effects of the Coulomb terms, energy level (ground or excited),
computational domains and numerical methodology on the convergence
of the solution to the two-electron problem.

\begin{table}
\caption{\label{DifferentRuns} A list of the different cases that are compared
in this section. Exc refers to the first excited {\it S} state. Grd refers to the ground state.
$N_D$ is the number of domains. B, R, and E refer to behavioral, regularity, and excision, respectively.
}
\begin{ruledtabular}
\begin{tabular}{|c|c|c|c|c|c|c|c|}
&\multicolumn{2}{c|}{Potential}&State&Domains&\multicolumn{3}{c|}{Boundary Conditions}\\
\hline
Case & $Z$& $\alpha$ & Exc/Grd & $N_D$ & $\rho=0$ & $\phi,\zeta=0$ & $\rho=\infty$\\
\hline
A&1&1&Grd&3&B&B&B\\
B&1&0&Grd&3&B&B&B\\
C&2&1&Grd&3&B&B&B\\
D&2&1&Exc&3&B&B&B\\
E&1&1&Grd&1&B&B&B\\
F&1&0&Grd&1&B&B&B\\
G&1&1&Grd&3&R&B&R\\
H&1&1&Grd&3&R&B&E\\
I&1&1&Grd&3&B&R&B\\
J&1&1&Grd&3&B&E&B
\end{tabular}
\end{ruledtabular}
\end{table}

\subsection{Convergence in energy}

In this section,
the energy error means the difference between the numerical
energy eigenvalue at finite resolution and the exact energy eigenvalue
of the nonrelativistic infinite-mass-nucleus Hamiltonian. When no
analytic value exists, highly precise variationally calculated values
are used
\cite{BakerEtAl1990,Drake1999,Korobov2000,Frolov2000,Korobov2002,DrakeEtAl2002,Frolov2006,
  Frolov2007,NakashimaNakatsuji2007,NakashimaNakatsuji2008,KurokawaEtAl2008}.
  
\begin{figure}
\includegraphics[width=\linewidth]{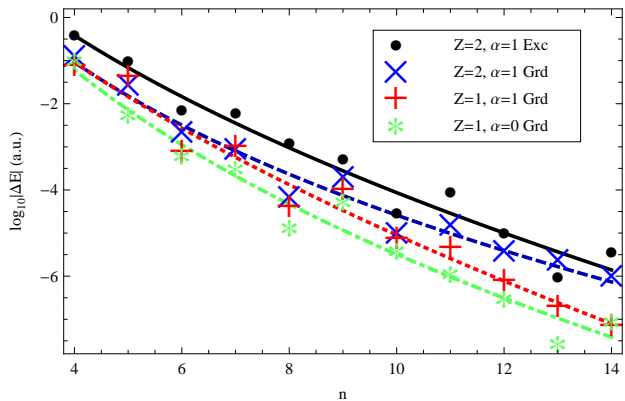}%
\caption{\label{EEOnOff} (Color online). The convergence of the energy of H$^-$, case A
  (red pluses); two non-interacting electrons in the field of a
  proton, case B (green stars); the ground state of He, case C (blue crosses); and its first excited
  S-state, case D (black circles) as a function of grid resolution $n$,
  with dotted red, dot-dashed green, dashed blue, and solid black fits, respectively. 
  See Appendix \ref{FitsSection} for fitting functions and method.}
\end{figure}

The energy errors for H$^-$, H$^-$ with the 
electron-electron interaction turned off,  the ground states of Helium, 
and the first excited {\it S} State 
of helium (cases A,B,C, and D) are
shown in Fig. \ref{EEOnOff}. The first
important result is that the energies appear to
converge in an approximate exponential fashion. 
Since these are not variational calculations there
is no reason to expect monotonically decreasing energy errors.
Detailed inspection of the solutions
suggests that the kinks in the graphs are discreteness
effects. That is, the precise positioning of the grid points has a
large effect on the magnitude of the error.

A potentially significant issue is the impact on convergence of
logarithmic terms present in the Fock expansion.  Prior authors have
been able to calculate very accurate energies without adverse effects
of the infinite second derivative at $\rho=0$, but the PS
method is sensitive to singularities anywhere within the
domain. Figure \ref{EEOnOff} includes a calculation of the H$^-$
system with the electron-electron interaction turned off, 
altering the exact solution and removing the logarithmic term. The
rate of convergence is comparable in all cases suggesting that the
influence of the logarithmic term on convergence is subdominant for
$n \le 14$. This conclusion agrees with the analysis in subsection \ref{OneDLogErr}.

Ideally, the numerical method should handle states
other than the ground state. Figure \ref{EEOnOff} shows the
convergence of energies for the ground state and the first excited {\it S} state of
helium. The important result is that the convergence of both
calculations is approximately exponential with a similar rate.

The relative sizes of the magnitude of the error at fixed grid size for
H$^-$, He and excited He are roughly consistent with the general
expectation set by the difficulty in resolving the
solution's small-scale structure. Errors for ground state He are
larger than H$^{-}$ because the exponential length scale for falloff of the
He wave function is smaller than that of H$^{-}$; errors for the
excited state of He are larger than the ground state of He because the
oscillatory length scale of the excited state is smaller than the 
exponential length scale of the ground state.


\begin{figure}
\includegraphics[width=\linewidth]{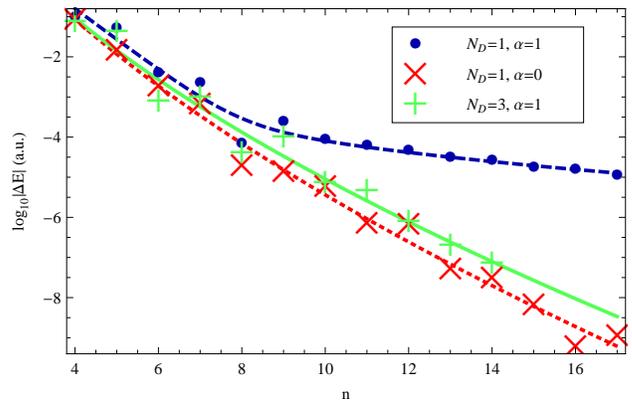}%
\caption{\label{DomComp} (Color online). The convergence of the energy of, case E, H$^-$ using
  only one computational domain with the electron-electron interaction
  on (blue circles); case F, one domain with the interaction off (red crosses); and
  case A, three domains with the interaction on (green pluses) with dashed blue,
  dotted red, and solid green fits, respectively. 
  See Appendix \ref{FitsSection} for fitting functions and method.}
\end{figure}

Figure \ref{DomComp} shows the impact on convergence of using a single
numerical subdomain, $D_0$, versus three, $\{D_1,D_2,D_3\}$. The
single domain had one third as many points as the computation with
three domains. However, the resolution in the $x$ direction dominates
the convergence and in that dimension the resolution is identical. 
Domain $D_0$ has radial-like coordinates
near the electron-proton cusp but not near the electron-electron
cusp. One anticipates slower convergence in the energy using
$D_0$. Comparison shows that two interacting electrons on three
domains (case A) or two noninteracting electrons on $D_0$ (case F) have similar
exponential rates of convergence. On the other hand, 
two interacting electrons on $D_0$ (case E) converge
more slowly. This result shows the multiple grids are essential for
achieving superior convergence, and that the electron-electron
cusp drives this requirement.
If the single domain data were adjusted to account for the fact
that they used fewer points, they would be shifted to the left
by a factor of $3^{1/3}\approx 1.44$. This would not affect the
conclusion that using three domains is more efficient because the 
single domain solution is only algebraically convergent starting at about 
$n=8$. Using three subdomains is more efficient because the work involved in the
calculation has the same scaling whether one or three subdomains is used.

\begin{figure}
\includegraphics[width=\linewidth]{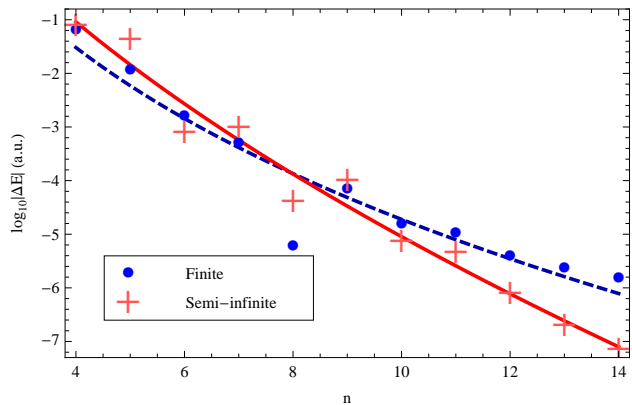}%
\caption{\label{RhoMax} (Color online). The convergence of the energy of, case H,  H$^-$ doing the
 calculation on a finite domain (blue circles); and, case A, 
semi-infinite domain (red pluses) with dashed blue and solid red fits, respectively. 
See Appendix \ref{FitsSection} for fitting functions and method.}
\end{figure}

Figure \ref{RhoMax} presents a comparison of calculations having the
full semi-infinite domain (case A) to those with a finite cutoff in $\rho$
(case H). The scaling of the cutoff $\rho_\textrm{max} \propto
\sqrt{n}$ imposed in case H is derived in Appendix \ref{Truncation} by
balancing the error due to finite resolution from the numerical scheme
with errors introduced by truncating the bound state. The figure shows that
the semi-infinite calculation fairs better. This is a consequence
of the two different sets of assumptions used to distribute the
points. The grid points in the semi-infinite scheme are more often
found where the wave function is large. Half the points have $\rho < 1$
($x>0$) because $0$ is the center of the $x$ dimension. By comparison,
half the points have $\rho<\rho_\textrm{max}/2$ in the finite calculation.
The number of points where the wave function is large is smaller
in this latter scheme.  Although the semi-infinite strategy is more effective,
nothing can be said about the optimal strategy because other
distribution methods were not considered. The main
advantage of the method
is simplicity since there are no adjustable parameters.

\begin{figure}
\includegraphics[width=\linewidth]{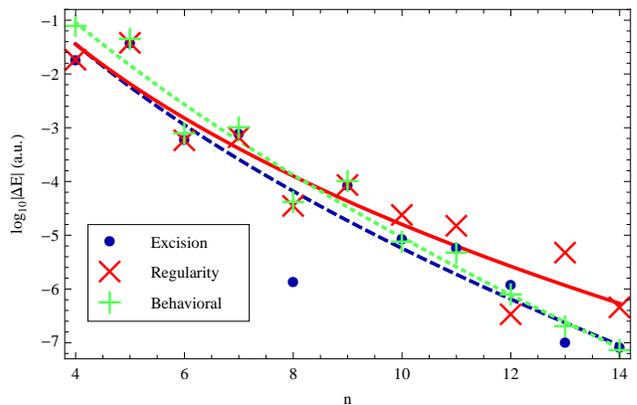}%
\caption{\label{CuspCond} (Color online). The convergence of the energy of H$^-$ using
  three different methods of ensuring regularity at the two-particle
  coalescence points: case A, relying on the regularity of the Chebyshev
  polynomials (green pluses); case I, using the Kato cusp condition as a
  regularity condition (red crosses); and case J, excising the singularity
  (blue circles) with green dotted, solid red, and dashed blue fits,
  respectively. See Appendix \ref{FitsSection} for fitting
  functions.}
\end{figure}

Figure \ref{CuspCond} presents a comparison of the different ways of
handling the regularity of the wave function at the two-particle
coalescence points.  The simplest method, relying on the
regularity of the Chebyshev polynomials (case A), does
as well or better than the other methods (cases I and J).

\begin{figure}
\includegraphics[width=\linewidth]{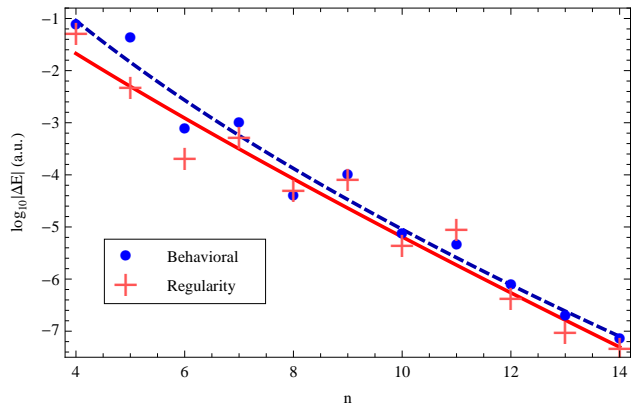}%
\caption{\label{TriplePt} (Color online). The convergence of the energy of H$^-$ using two different methods of ensuring regularity at the 
three-particle coalescence point: case A, relying on the regularity of the Chebyshev polynomials (blue circles); and case G,
using the Fock condition 
to specify a logarithmic derivative (red pluses) 
with dashed blue and solid red fits, respectively. See Appendix \ref{FitsSection} for fitting functions and method.}
\end{figure}

Figure \ref{TriplePt} compares two ways of handling the
wave function at $\rho=0$, case A, relying on the regularity of the Chebyshev
polynomials (behavioral) and case G, directly specifying a logarithmic
derivative (regularity).  The latter method is slightly better but
both have roughly the same convergence rate. 

\subsection{Convergence in local energy}

\begin{figure}
\includegraphics[width=\linewidth]{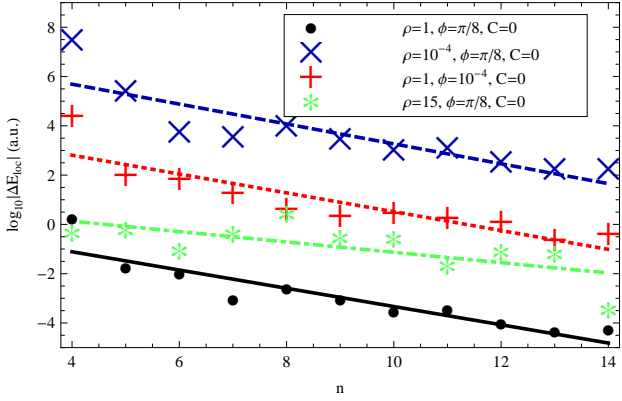}%
\caption{\label{ELocn} (Color online). The convergence of the local energy of H$^-$ at
  a four points in the domain: the center of the computational domain (black circles),
  near the triple coalescence point (blue crosses), near the proton-electron coalescence
  point (red pluses), and at large $\rho$ (green stars). Their geometric fits are 
  given by the solid black, dashed blue, dotted red, and dot-dashed green lines, 
  respectively. See Appendix \ref{FitsSection} for fitting functions and method.
 }
\end{figure}

Another useful measure of convergence is the local energy,
\begin{equation}
E_{\textrm{loc}}=\frac{\cal{H}\psi}{\psi}.
\end{equation}
which is constant only for an exact eigenfunction $\psi$ of
Hamiltonian $\cal H$.
Throughout this subsection all analysis and data refers to case A.

The difference between the local energy and the numerically
evaluated eigenvalue $E$ gives a local measure of the error 
in $\psi$ in a particular calculation. Define
\begin{equation}
\Delta E_{\textrm{loc}}=E_{\textrm{loc}}-E.
\end{equation}
For the PS method, $\Delta E_{\textrm{loc}}$ is zero at
all grid points (subject to limits of finite precision arithmetic).
Nonzero differences exist between grid points.  Figure \ref{ELocn}
illustrates the convergence of local energy at four different points. 
Of the four points, the error in local energy is lowest at the point 
in the center of the computational domain (\mbox{$\{\rho,\phi,C\}=\{1,\pi/8,0\}$}).
It is larger in magnitude near the singularities 
(\mbox{$\{\rho,\phi,C\}=\{10^{-4},\pi/8,0\}$} and \mbox{$\{1,10^{-4},0\}$})
because near these points 
$E_{\textrm{loc}}\rightarrow\infty$ for any nonexact $\psi$.
However, the  geometric fits show that the rate of convergence is approximately the
same at all three of those points. A different behavior is seen  
at \mbox{$\{\rho,\phi,C\}=\{15,\pi/8,0\}$}. The error 
is roughly constant over much of the graph, but begins to 
decrease at high resolution. This is not surprising given the
fact that there are only a few grid points at such large hyperradius.

\begin{figure}
\includegraphics[width=\linewidth]{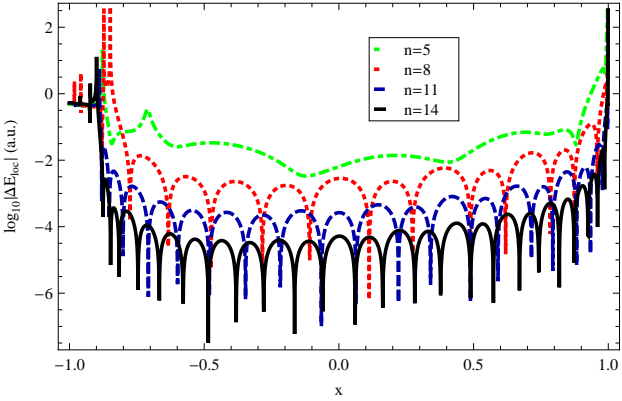}%
\caption{\label{ELocRes} (Color online). The error in the local energy of H$^-$ as a function of $x$ with
$r_1=2r_2$, and $C=-1$ (the electrons are on the same side of the nucleus) at four different resolutions:
$n=5$ green dot-dashed, $n=8$ red dotted, $n=11$ blue dashed, $n=14$ black solid.}
\end{figure}

Figure \ref{ELocRes} displays the convergence of the local energy as a
function of $x$ at fixed angular coordinates. For a perfect exponential
decrease in local energy error,
the curves would be equidistant from each other. This is approximately true 
throughout the domain except near $x=\pm 1$ (small and large $\rho$).

If the numerical solution is considered to be trustworthy where the local 
energy error is less than some threshold (e.g.
$|\Delta E_{\textrm{loc}}| < 10^{-2}$), then the
wave function is well represented in an
intermediate range of $\rho$ ($10^{-2}<\rho<10^{1.3}$) but
not near the triple coalescence point nor at infinity
\footnote{The error in the function value is roughly constant everywhere 
in the domain, but the magnitude of derivatives and the wave function becomes large compared to 
the magnitude of the wave function at small and large $\rho$.}.

At large $\rho$ ($x\approx -1$), the true wave function falls off exponentially
(in fact, with respect to the $x$ coordinate it falls off even faster).
The PS method represents the exponential in terms of
a polynomial. When one extrapolates using the polynomial to $x=-1$,
the wave function is small but nonzero (the exact value should be zero).
However, the Hamiltonian acting on the polynomial is guaranteed to be zero
because every coefficient in the Hamiltonian operator has a factor of $(1+x)$.
So 
\begin{equation}
\lim_{\rho\rightarrow\infty} \frac{{\cal{H}}\psi}{\psi}=0,
\end{equation}
and $\Delta E_{\textrm{loc}} \to -E$, 
a constant at large $\rho$ as seen in Fig. \ref{ELocRes}.
Detailed inspection of the data near $x=-1$ suggests that for any finite $\rho$ there exists a resolution above
which the solution becomes trustworthy. 

\begin{figure}
\includegraphics[width=\linewidth]{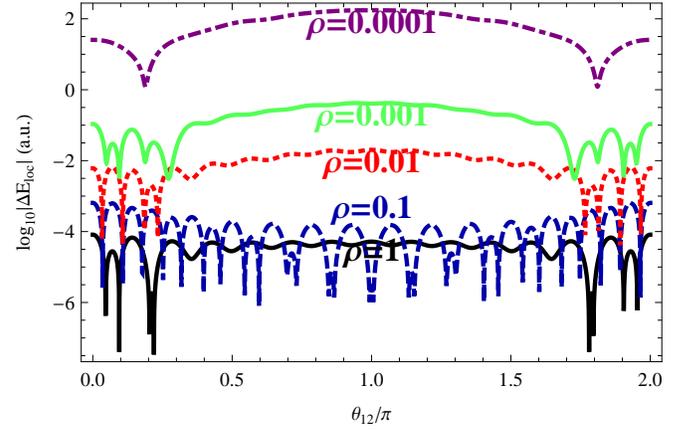}%
\caption{\label{LocEnErrRho} (Color online). The error in local energy of H$^-$ plotted at different values of the angle
$\theta_{12}$ with $r_1=2r_2$ and at resolution $n=14$ and at $\rho=1$ (solid black), $\rho=0.1$ (dashed blue),
$\rho=0.01$ (dotted red), $\rho=0.001$ (solid greed), and $\rho=0.0001$ (dot-dashed purple).}
\end{figure}


The local energy behavior near the triple-coalescence point ($\rho=0$)
is of special interest as a probe of the wave function's nonanalytic
behavior. Figure \ref{LocEnErrRho} displays $\Delta E_{\textrm{loc}}$
as a function of $\theta_{12}$, the angle between the two electrons,
for fixed $r_2/r_1$ and for a number of choices of $\rho$, following a
similar figure from Myers {\it et al} \cite{MyersEtAl1991}. 
In this small $\rho$ regime, the terms that dominate the Hamiltonian are the 
kinetic energies in the various directions. Each of these, individually, scales as
$1/\rho^2$ [see Eq. (\ref{Hamiltonian})]. For the exact solution, these terms 
cancel each other, but for almost any solution
which is not exact, the local energy scales as $1/\rho^2$. This scaling is 
shown in Figs. \ref{ELocRes} and \ref{LocEnErrRho}. 
Again detailed inspection of the data near $x=1$ suggests that for any finite 
nonzero $\rho$ there exists a resolution above
which the solution becomes trustworthy. 
Furthermore, Fig. \ref{ELocn} shows that there is no sign of the convergence
rate being slowed due to the logarithmic terms at this resolution.

\subsection{Cauchy errors}

\begin{figure}
\includegraphics[width=\linewidth]{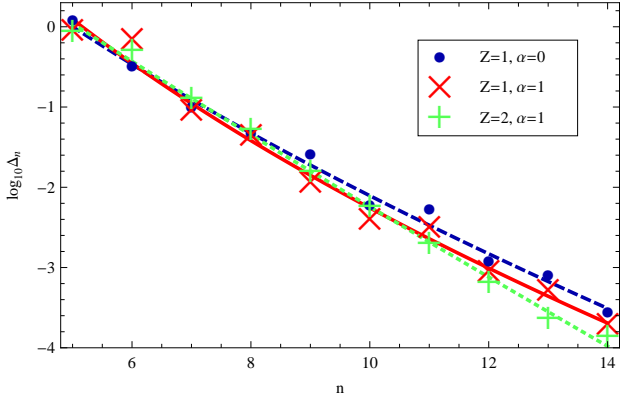}%
\caption{\label{CauchyConv} (Color online). The root-mean-square average Cauchy error is plotted with increasing resolution
for three cases: H$^-$ with noninteracting electrons (blue circles); H$^-$ with interacting electrons 
(red crosses); and helium with interacting electrons (green pluses) with dashed blue, solid red, and 
dotted green fits, respectively. See Appendix \ref{FitsSection} for fitting functions and method.}
\end{figure}

Throughout this subsection all data refers to cases A, B, and C.
The Cauchy error is a measure of the 
difference between numerical solutions
with different resolution. One such measure is the normed quantity
\begin{equation}
\Delta_n=\sqrt{\int \ud^3r_1\,\ud^3r_2\,(\psi_{n}-\psi_{n-1})^2}.
\end{equation}
The true $\psi$ satisfies
\begin{equation}
1=\int \ud^3r_1\,\ud^3r_2\,\psi^2,
\end{equation}
but integrating $\psi_n$ all the way to $\rho=\infty$ would diverge.
This is a consequence of having small but nonzero errors at $\rho=\infty$ in 
the value of $\psi_n$.  
An upper limit $\rho=10$ is adopted in the normalization of $\psi_n$
and calculation of $\Delta_n$. It is arbitrary but encompasses most of the physical
extent of the solution. 
The Cauchy error in any subinterval of the
full interval must converge. To the extent that the
error in the interval calculated is dominant, the rate of
convergence can be assessed.

\begin{figure}
\includegraphics[width=\linewidth]{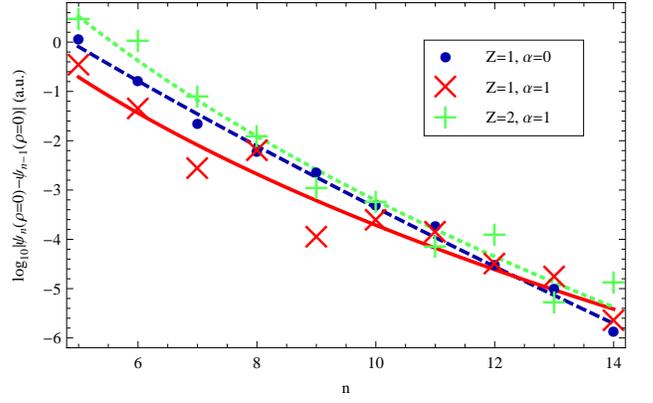}%
\caption{\label{CauchyNInf} (Color online). Pointwise differences in the wave function
  evaluated at $\rho=0$ for increasing resolution for three cases:
  H$^-$ with noninteracting electrons (blue circles); H$^-$ with
  interacting electrons (red crosses); and helium with interacting
  electrons (green pluses) with dashed blue, solid red, and 
dotted green fits, respectively. See Appendix \ref{FitsSection} for fitting functions and method.}
\end{figure}

Figure \ref{CauchyConv} gives $\Delta_n$ as a function of resolution
while Fig. \ref{CauchyNInf} gives the pointwise difference 
at $\rho=0$ where the wave function is maximum. Both
plots show that convergence is approximately exponential.

\subsection{The logarithmic derivative at the triple coalescence point}

\begin{figure}
\includegraphics[width=\linewidth]{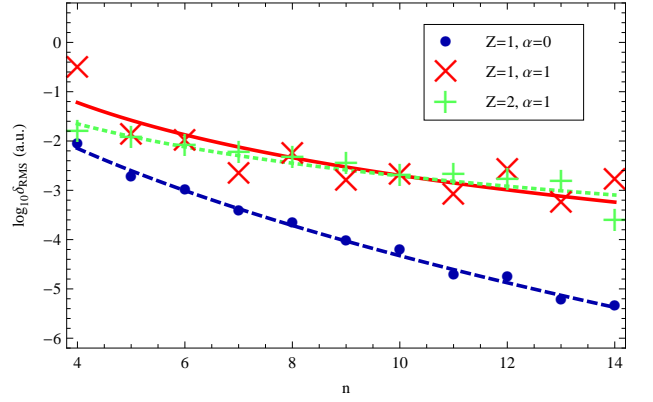}%
\caption{\label{LogDer} (Color online). The root-mean-square error in the logarithmic
  derivative evaluated at $\rho=0$ with increasing resolution for
  three cases: H$^-$ with noninteracting electrons (blue circles); $H^-$
  with interacting electrons (red crosses); and helium with interacting
  electrons (green pluses) with dashed blue, solid red, and 
dotted green fits, respectively. See Appendix \ref{FitsSection} for fitting functions and method.}
\end{figure}

Throughout this subsection all data refers to cases A, B, and C.
The only direct evidence
that the convergence of the solutions 
is slowed by the logarithmic terms in the exact solution comes from
evaluating the logarithmic derivative with respect to $\rho$ at $\rho=0$. 
The exact value is
\begin{equation}
\left.\frac{\partial_x\psi}{\psi}\right|_{x=1}
=-\frac{1}{2}\left(-Z(\cos\phi+\sin\phi)+\frac{\alpha}{2}\sigma[C,\phi]\right).
\end{equation}
The root-mean-square error is
\begin{equation}
\delta_{\textrm{RMS}}=\sqrt{\int \ud\Omega 
\left(\frac{\partial_x\psi}{\psi}-\frac{\partial_x\psi_n}{\psi_n}\right)_{x=1}^2},
\end{equation}
where
\begin{equation}
\int \ud\Omega =\int_0^{\pi/4}\ud\phi\, \sin^2 2\phi \int_{-1}^1 \ud C.
\end{equation}
Figure \ref{LogDer} displays $\delta_\textrm{RMS}$. Turning off the
electron-electron interaction, the convergence is noticeably
faster.  The important conclusion is that the
wave functions' convergence is indistinguishable from exponentially
fast, while the convergence of its derivatives, at least near $\rho=0$
is slower. Of course, the second derivative with respect to $\rho$ is
infinitely wrong at $\rho=0$. It converges to a finite value, and the 
exact value is infinite.

\section{Conclusion}

This article demonstrates the application of
PS methods for solving the nonrelativistic
Schr\"{o}dinger equation for a system with two electrons.  The method
successfully handled both ground and excited {\it S} states of heliumlike
systems.

The rate of convergence for most properties measured was
indistinguishable from being exponentially fast.  Local errors 
decrease in the same manner.

The choice of variables in the vicinity of the two-particle
coalescence and the use of multiple, overlapping domains are the
critical requirements. These are important so the PS method can
represent the analytic form of the solution near all the two-particle
cusps and ensure a more efficient algorithm. In other respects the most straightforward choices
work well. For example, grid points are determined by the roots of
Chebyshev polynomials, which experience shows generally produce the
best convergence in PS methods \cite{Fornberg1996,Boyd2000}. 
Behavioral boundary conditions (no explicit regularity
conditions) are sufficient to handle the wave function in the vicinity
of the coalescence points and also produce 
convergence as good as or better than the other possibilities tested.

The energy eigenvalue found by the variational method converges most
efficiently when basis functions which behave like the exact solution
are included but this selection process can be time-consuming and
problematic. Of course, much higher precision than reported here was obtained long ago by
variational methods. The PS method has
the advantage in new and possibly also in more complex applications of
not needing the same sort of specialized tuning that has benefited
variational calculations. Although this article does not attempt to reproduce the
ultra-high-precision results achieved by variational methods it strongly
suggests that the PS method will ultimately prove
to be a superior approach for reaching such results in
systems with a small number of electrons.

The local energy is not directly controlled when total energy is
minimized.  Local energy minimization schemes exist and
have the advantage that excited states are found at local minima of the
variance in local energy instead of just the ground state as is true
for the standard variational method \cite{UmrigarEtAl1988}. 
However, they lead to nonlinear problems which may be difficult to
handle numerically (minimization of the variance in local energy with 
respect to parameters in the trial wave function), but still tractable
because one need not calculate the energy at each step. 
By contrast, the PS method controls local energy while the
numerical solution remains a linear one.  At the same time, the
PS method is superior in terms of its convergence rate to
other direct partial differential equation solvers (grid based
methods) such as finite differencing and finite element which also
control the local error.


We plan to extend the method to calculate non-{S\it} states and continuum
two-electron states to compute the photoabsorption bound-free cross
sections with both initial and final states evaluated with the same
methodology.

\appendix
\section{Setting $\rho_\textrm{max}$ for the Truncated Domain}
\label{Truncation}

\begin{figure}
\includegraphics[width=\linewidth]{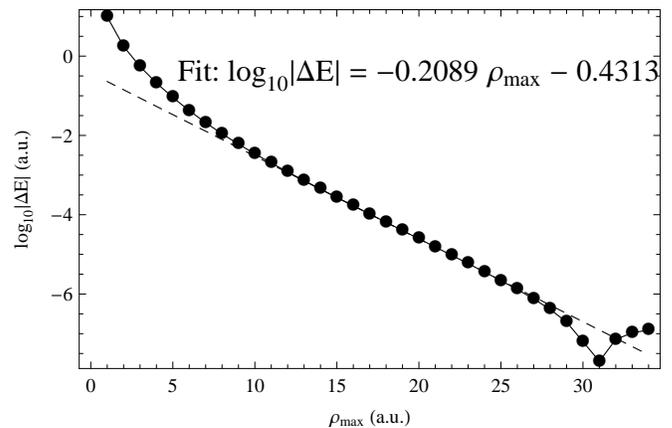}%
\caption{\label{EvsR} Energy error from truncation as a function of $\rho_\textrm{max}$
with a fit to the points from $\rho_\textrm{max}=15$ to $\rho_\textrm{max}=25$.}
\end{figure}

\begin{figure}
\includegraphics[width=\linewidth]{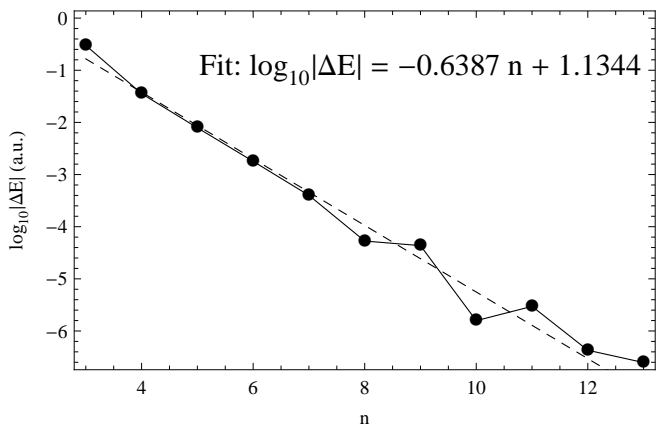}%
\caption{\label{EvsN} Energy error from finite resolution 
as a function of resolution at $\rho_\textrm{max}=20$
with a fit.}
\end{figure}

When truncating the domain at a finite $\rho$, it is wise to balance
  the error produced by the cutoff with that given by the finite
  resolution. The former was studied by calculating the energy of
  case H as a function of $\rho_\textrm{max}$. Figure \ref{EvsR} shows that difference 
  between the truncated energy and the correct value of the H$^-$ energy
  falls off as an exponential. The calculation was done with $n=14$. 
  We see at large values of $\rho_\textrm{max}$ that this finite resolution
  ruins the exponential behavior. There is a minimum at $\rho_\textrm{max}=31$.
  This is probably where the resolution error happens to cancel the truncation
  error. At larger $\rho_\textrm{max}$, the resolution error dominates. Therefore,
  only the points with $15\le\rho_\textrm{max}\le 25$ were used for the fit,
  $\log_{10}|\Delta E|=A \rho_\textrm{max} + B$. $A$ and $B$ were found to be 
  $-0.2089$ and $-0.4313$, respectively.
 
In order to measure the effect of finite resolution $\rho_\textrm{max}$ was
fixed at 20 and the difference between the energy at resolution $n$ and $14$
was plotted in Fig. \ref{EvsN}. The error from resolution effects should
increase when $\rho_\textrm{max}$ is increased because the density of points 
goes down. So the resolution error is assumed to be of the form 
$\log_{10}|\Delta E|=C n/\rho_\textrm{max} + D$. $C/20$ and $D$ were found to be
$-0.6387$ and $1.1344$, respectively.
  
Setting the two errors equal to each other yields the formula
\begin{equation}
\rho_\textrm{max} = -3.7476 + 7.8100 \sqrt{n+0.2297}.
\end{equation}

It is technically possible to get better energies as was shown in 
Fig. \ref{EvsR} (due to error cancellations), but taking advantage 
of that kind of effect is fine-tuning.

\section{\label{FitsSection} Fits}

\begin{turnpage}
\begin{table*}
\caption{\label{FitsTable} The convergence fits of quantities, $Q$, displayed in figures throughout the article.}
\begin{ruledtabular}
\begin{tabular}{ccccccccccc}
$Q$ & Figure(s) & Fitting Function & $a_1$ & $\Delta a_1$ & $a_2$ & $\Delta a_2$ & $a_3$ & $\Delta a_3$ & $a_4$ & $\Delta a_4$\\
\hline
$\delta_\textrm{RMS}^N[g_1]$ & \ref{CuspErr} & $a_1 n^{a_2}$ & $0.470$ & $0.189$ & $-1.852$ & $0.206$ &&&&\\
$\delta_\textrm{RMS}^N[g_1]$ & \ref{CuspErr} & $a_1 (a_2)^{n^{a_3}}$ & $0.895$ & $0.070$ & $0.4031$ & $0.0015$ & $1.3846$ & $0.0017$ &&\\
$|\Delta f[\rho=10^{-6}]|$ & \ref{LogErr} & $a_1 (a_2)^n + a_3 n^{a_4}$ & $0.0818$ & $0.0487$ & $0.696$ & $0.022$ & $0.295$ & $0.113$ & $-3.822$ & $0.092$ \\
$|\Delta f[\rho=10^{-1}]|$ & \ref{LogErr} & $a_1 (a_2)^n + a_3 n^{a_4}$ & $0.0264$ & $0.0306$ & $0.725$ & $0.027$ & $0.472$ & $0.595$ & $-5.172$ & $0.290$ \\
$|\Delta f[\rho=10^{4}]|$ & \ref{LogErr} & $a_1 (a_2)^n + a_3 n^{a_4}$ & $0.161$ & $0.112$ & $0.743$ & $0.011$ & $1.29\times 10^{-7}$ & $2.16\times 10^{-7}$ & $-2.460$ & $0.372$ \\
$|\Delta H_N^\textrm{max}|(\alpha=1)$ & \ref{DHMax} & $a_1 (a_2)^{n^{a_3}}$ & $936.$ & $418.$ & $0.5943$ & $0.0089$ & $0.8348$ & $0.0067$ &&\\
$|\Delta H_N^\textrm{max}|(\alpha=0)$ & \ref{DHMax} & $a_1 (a_2)^{n^{a_3}}$ & $14.6$ & $6.0$ & $0.3282$ & $0.0073$ & $0.7209$ & $0.0047$ &&\\
$|(\Delta H_N)^{i^*}|(\alpha=1)$ & \ref{DHIn} & $a_1 (a_2)^n + a_3 n^{a_4}$  & $1.96\times 10^{-7}$ & $1.36\times 10^{-7}$ & $0.491$ & $0.018$ & $0.263$ & $0.091$ & $-10.358$ & $0.083$ \\
$|(\Delta H_N)^{i^*}|(\alpha=0)$ & \ref{DHIn} & $a_1 (a_2)^{n^{a_3}}$  & $181.$ & $84.$ & $2.90\times 10^{-6}$ & $3.1\times 10^{-7}$ & $0.3733$ & $0.0020$ \\
$|\Delta E|$ (Case A) & \ref{EEOnOff},\ref{DomComp},\ref{RhoMax},\ref{CuspCond},\ref{TriplePt} & $a_1 (a_2)^{n^{a_3}}$ & $121000.$ & $64000.$ & $0.00138$ & $0.00022$ & $0.549$ & $0.01$ &&\\
$|\Delta E|$ (Case B) & \ref{EEOnOff} & $a_1 (a_2)^{n^{a_3}}$ & $1.17\times 10^{11}$ & $6.7\times 10^{10}$ & $1.53\times 10^{-8}$ & $4.3\times 10^{-9}$ & $0.3262$ & $0.0070$ &&\\
$|\Delta E|$ (Case C) & \ref{EEOnOff} & $a_1 (a_2)^{n^{a_3}}$ & $1.28\times 10^{11}$ & $5.7\times 10^{10}$ & $5.71\times 10^{-9}$ & $1.39\times 10^{-9}$ & $0.2796$ & $0.0057$ &&\\
$|\Delta E|$ (Case D) & \ref{EEOnOff} & $a_1 (a_2)^{n^{a_3}}$ & $1.62\times 10^{6}$ & $880000.$ & $0.000385$ & $0.000073$ & $0.478$ & $0.011$ &&\\
$|\Delta E|$ (Case E) & \ref{DomComp} & $a_1 (a_2)^n + a_3 n^{a_4}$ & $193.$ & $140.$ & $0.169$ & $0.021$ & $0.190$ & $0.087$ & $-3.39$ & $0.18$ \\
$|\Delta E|$ (Case F) & \ref{DomComp} & $a_1 (a_2)^{n^{a_3}}$ & $286000.$ & $126000.$ & $0.00100$ & $0.00012$ & $0.5604$ & $0.0069$ &&\\
$|\Delta E|$ (Case G) & \ref{TriplePt} & $a_1 (a_2)^{n^{a_3}}$ & $30.0$ & $19.3$ & $0.0972$ & $0.0100$ & $0.819$ & $0.019$ &&\\
$|\Delta E|$ (Case H) & \ref{RhoMax} & $a_1 (a_2)^{n^{a_3}}$ & $2.10\times 10^{10}$ & $1.52\times 10^{10}$ & $5.73\times 10^{-9}$ & $2.36\times 10^{-9}$ & $0.2614$ & $0.0097$ &&\\
$|\Delta E|$ (Case I) & \ref{CuspCond} & $a_1 (a_2)^{n^{a_3}}$ & $3.89\times 10^{9}$ & $3.10\times 10^{9}$ & $4.08\times 10^{-8}$ & $1.73\times 10^{-8}$ & $0.289$ & $0.011$ &&\\
$|\Delta E|$ (Case J) & \ref{CuspCond} & $a_1 (a_2)^{n^{a_3}}$ & $5.26\times 10^{6}$ & $5.68\times 10^{6}$ & $0.0000263$ & $0.0000114$ & $0.417$ & $0.018$ &&\\
$|\Delta E_\textrm{loc}[\rho=1,\phi=\pi/8,C=0]|$ & \ref{ELocn} & $a_1 (a_2)^{n}$ & $2.35$ & $1.97$ & $0.426$ & $0.037$ &&&&\\
$|\Delta E_\textrm{loc}[\rho=10^{-4},\phi=\pi/8,C=0]|$ & \ref{ELocn} & $a_1 (a_2)^{n}$  & $2.02\times 10^{7}$ & $2.42\times 10^{7}$ & $0.395$ & $0.050$ &&&&\\
$|\Delta E_\textrm{loc}[\rho=1,\phi=10^{-4},C=0]|$ & \ref{ELocn} & $a_1 (a_2)^{n}$  & $21700.$ & $21900.$ & $0.415$ & $0.044$ &&&&\\
$|\Delta E_\textrm{loc}[\rho=15,\phi=\pi/8,C=0]|$ & \ref{ELocn} & $a_1 (a_2)^{n}$  & $9.33$ & $10.73$ & $0.616$ & $0.074$ &&&&\\
$\Delta_n(Z=1,\alpha=0)$ & \ref{CauchyConv} & $a_1 (a_2)^{n^{a_3}}$ & $7260.$ & $1200.$ & $0.0394$ & $0.0016$ & $0.6282$ & $0.0053$ &&\\
$\Delta_n(Z=1,\alpha=1)$ & \ref{CauchyConv} & $a_1 (a_2)^{n^{a_3}}$ & $2.77\times 10^{6}$ & $590000.$ & $0.000876$ & $0.000068$ & $0.4541$ & $0.0047$ &&\\
$\Delta_n(Z=2,\alpha=1)$ & \ref{CauchyConv} & $a_1 (a_2)^{n^{a_3}}$ & $472.$ & $58.$ & $0.2406$ & $0.0038$ & $0.9001$ & $0.0047$ &&\\
$|\psi_n(\rho=0)-\psi_{n-1}(\rho=0)|(Z=1,\alpha=0)$ & \ref{CauchyNInf} & $a_1 (a_2)^{n^{a_3}}$ & $27300.$ & $5900.$ & $0.0520$ & $0.0019$ & $0.7833$ & $0.0052$ &&\\
$|\psi_n(\rho=0)-\psi_{n-1}(\rho=0)|(Z=1,\alpha=1)$ & \ref{CauchyNInf} & $a_1 (a_2)^{n^{a_3}}$ & $1.17\times 10^{12}$ & $6.9\times 10^{11}$ & $1.49\times 10^{-8}$ & $4.5\times 10^{-9}$ & $0.3046$ & $0.0073$ &&\\
$|\psi_n(\rho=0)-\psi_{n-1}(\rho=0)|(Z=2,\alpha=1)$ & \ref{CauchyNInf} & $a_1 (a_2)^{n^{a_3}}$ & $1.63\times 10^{14}$ & $8.6\times 10^{13}$ & $1.49\times 10^{-8}$ & $3.6\times 10^{-9}$ & $0.3475$ & $0.0059$ &&\\
$\delta_\textrm{RMS}(Z=1,\alpha=0)$ & \ref{LogDer} & $a_1 (a_2)^{n^{a_3}}$ & $170.$ & $23.$ & $0.00418$ & $0.00022$ & $0.4404$ & $0.0041$ &&\\
$\delta_\textrm{RMS}(Z=1,\alpha=1)$ & \ref{LogDer} & $a_1 n^{a_2}$  & $10.3$ & $6.1$ & $-3.71$ & $0.27$ &&&&\\
$\delta_\textrm{RMS}(Z=2,\alpha=1)$ & \ref{LogDer} & $a_1 n^{a_2}$ & $0.883$ & $0.275$ & $-2.65$ & $0.14$ &&&& 

\end{tabular}
\end{ruledtabular}
\end{table*}
\end{turnpage}

Numerical rates for convergence are summarized in Table
\ref{FitsTable}. 
The method is as follows: let $Q_n$ be the quantity
converging to zero as $n$ goes to infinity, consider fitting functions
of the forms
\begin{eqnarray}
f_\textrm{geom}[n] & = & a_1 (a_2)^n\\
f_\textrm{alg}[n] & = & a_1 n^{a_2}\\
f_\textrm{sup/sub}[n]&=&a_1 (a_2)^{n^{a_3}}\\
f_\textrm{geom/alg}[n]&=&a_1 (a_2)^n + a_3 n^{a_4},
\end{eqnarray}
which are geometric, algebraic, supergeometric ($a_3>1$) or subgeometric 
($a_3<1$), and mixed geometric and algebraic fits, respectively. 
A $\chi^2$ fitting method is used:
\begin{equation}
\chi^2  =  \sum_n \left(\log_{10} \left|\frac{Q_n}{f[n]}\right|\right)^2,
\end{equation}
where $Q_n$ is the quantity $Q$ evaluated at resolution $n$ and the
sum over $n$ goes from the even numbers from $8$ to $100$ for the 
one-dimensional models, $5$ to $14$ for the Cauchy errors, and from $4$ to $14$
for all others. $\chi^2$ is minimized with respect to all $a_i$ for
the fit which is most reasonable on theoretical grounds. Errors in 
the $a_i$ are estimated by calculating
\begin{equation}
\Delta a_i = \sqrt{\frac{\chi^2}{\partial_{a_i a_i} \chi^2}}
\end{equation}
at the minimum of $\chi^2$. Of course large error in the values of amplitudes
$a_1$ and $a_3$ for the mixed geometric and algebraic fit imply
that the other parameters that multiply that amplitude may be meaningless.

\begin{acknowledgments}
We thank Saul Teukolsky, Harald Pfeiffer, Cyrus Umrigar, Ira Wasserman,
 Sharvari Nadkarni-Ghosh, and Sergei Dyda for helpful conversations. 
This material is based on work supported by the National
        Science Foundation under Grant No. AST-0406635.
\end{acknowledgments}

\bibliography{HMinPRA}

\end{document}